\documentclass[journal]{IEEEtran}
\IEEEoverridecommandlockouts
\usepackage{amsfonts}
\usepackage{amsmath}
\usepackage{amssymb}
\usepackage{algorithm}
\usepackage{algorithmicx}
\usepackage{algpseudocode}
\usepackage{dsfont}
\usepackage{authblk}
\usepackage{bbm}
\usepackage{graphicx}
\usepackage{epstopdf}
\usepackage{subfigure}
\usepackage{stfloats}
\usepackage{eqnarray}
\usepackage{makecell}
\usepackage{multirow}
\usepackage{enumerate}
\usepackage{booktabs}
\usepackage{cite}
\usepackage{balance}
\usepackage{color}
\usepackage{bm}
\usepackage{caption}
\usepackage{mathtools}
\DeclarePairedDelimiter\autobracket{(}{)}
\DeclarePairedDelimiter\autosquare{[}{]}
\newcommand{\br}[1]{\autobracket*{#1}}
\newcommand{\sq}[1]{\autosquare*{#1}}
\providecommand{\OO}[1]{\mathcal{O}\left(#1\right)} 

\newcommand{\argmin}{\rm argmin}
\newcommand{\argmax}{\rm argmax}

\renewcommand{\IEEEQED}{\IEEEQEDopen}

\newcommand{\Tr}{\mbox{Tr}}

\newcommand{\REQ}{\mbox{\tiny REQ}}
\newcommand{\TOT}{\mbox{\tiny TOT}}
\newcommand{\SEC}{\mbox{\tiny SEC}}

\newcommand{\OPT}{\cal OPT}
\newcommand{\SINR}{\mbox{SINR}}
\renewcommand{\vec}{\mbox{vec}}
\renewcommand{\H}{\mbox{\tiny H}}
\newcommand{\F}{\mbox{\tiny F}}
\newcommand{\Rank}{\mbox{Rank}}
\newcommand{\C}{\mbox{\tiny CIR}}
\newcommand{\Q}{\bm Q}
\newcommand{\W}{\bm W}
\newcommand{\st}{\mbox{s.t.}}
\newcommand{\sumn}{\sum\limits_{n=1}^N}
\newcommand{\sumnk}{\sum\limits_{k=1, k \neq n}^N}
\newcommand{\sumk}{\sum\limits_{k=1}^N}

\newcommand{\diag}{\mbox{diag}}

\newtheorem{lemma}{\textbf{Lemma}}

\newtheorem{remark}{\textbf{Remark}}
\newtheorem{proposition}{\textbf{Proposition}}

\begin{document}
\title{Robust Energy Efficient Beamforming in MISOME-SWIPT Systems With Proportional Secrecy Rate}

\author{
\mbox{Yanjie Dong,~\IEEEmembership{Student Member, IEEE},}
\mbox{Md. Jahangir Hossain,~\IEEEmembership{Senior Member, IEEE},}
\mbox{Julian Cheng,~\IEEEmembership{Senior Member, IEEE},}
\mbox{and Victor C. M. Leung,~\IEEEmembership{Fellow, IEEE}}
\thanks{
This paper was accepted in part by the IEEE Globecom, Abu Dhabi, UAE, Dec. 9--13, 2018 \cite{Dong2018b}.
This work was supported in part by the National Natural Science Foundation of China under Grant 61671088, in part by a UBC Four-Year Doctoral Fellowship, and in part by the Natural Science and Engineering Research Council of Canada. (\emph{Corresponding author: Victor C. M. Leung.})}
\thanks{Y. Dong is with the Department of Electrical and Computer Engineering, The University of British Columbia, Vancouver, BC, V6T 1Z4 Canada (email:ydong16@ece.ubc.ca).}
\thanks{V. C. M. Leung is with the Department of Electrical and Computer Engineering, The University of British Columbia, Vancouver, BC V6T 1Z4, Canada, and also with Zhejiang Gongshang University, Hangzhou 310018, China (e-mail: vleung@ece.ubc.ca).}
\thanks{M. J. Hossain and J. Cheng are with the School of Engineering, The University of British Columbia, Kelowna, BC, V1V 1V7 Canada (email:\{jahangir.hossain, julian.cheng\}@ubc.ca).}
}

\maketitle
\begin{abstract}
The joint design of beamforming vector and artificial noise covariance matrix is investigated for the multiple-input-single-output-multiple-eavesdropper simultaneous wireless information and power transferring \mbox{(MISOME-SWIPT)} systems.
In the MISOME-SWIPT system, the base station delivers information signals to the legitimate user equipments and broadcasts jamming signals to the eavesdroppers.
A secrecy energy efficiency (SEE) maximization problem is formulated for the considered \mbox{MISOME-SWIPT} system with imperfect channel state information, where the SEE is defined as the ratio of sum secrecy rate over total power consumption.
Since the formulated SEE maximization problem is non-convex, it is first recast into a series of convex problems in order to obtain the optimal solution with a reasonable computational complexity.
Two suboptimal solutions are also proposed based on the heuristic beamforming techniques that trade performance for computational complexity.
In addition, the analysis of computational complexity is performed for the optimal and suboptimal solutions.
Numerical results are used to verify the performance of  proposed algorithms and to reveal practical insights.
\end{abstract}
\begin{IEEEkeywords}
Beamforming, energy efficient communications, MISOME-SWIPT, proportional secrecy rate, secrecy energy efficiency maximization.
\end{IEEEkeywords}

\section{Introduction}
Energy harvesting technology, which enables devices to scavenge energy from ambient resources, has attracted significant attention from academia and industry in recent decade \cite{Lu2015, HasanFourthquarter2011}.
Harvesting energy from natural resources requires the natural-resource energy-harvesting (NR-EH) module to be directly exposed in natural resources, such as solar and wind \cite{HasanFourthquarter2011, DongDec.2017, Guotobepublished}.
Moreover, the NR-EH module  has a large form factor \cite{HasanFourthquarter2011}, which is not suitable for small devices, e.g., implantable wireless sensors.
Another drawback of NR-EH takes root in the volatility of natural resources.
Since the radio-frequency (RF) signals deliver both information and energy as well as penetrate the obstacles, radio-frequency energy-harvesting (RF-EH) technology becomes an alternative to overcome the shortcomings of NR-EH technology.
Since the energy of RF signals is controllable, the RF-EH technology is more suitable for stable energy provision \cite{Dong2016, BiApri2015, ZengMay}.

The current research on RF-EH technology can be categorized into two branches.
The first branch focuses on wireless powered communication systems (WPCSs), where the wireless devices harvest energy from RF signals and then use the harvested energy to transmit information \cite{ZengMay, JuJan.2014, JuOct.2014}.
For example, the authors in \cite{JuJan.2014, JuOct.2014} proposed algorithms to maximize the system throughput in the multiuser WPCSs, where the base station (BST) broadcasts the energy signals and receives information signals in half-duplex mode \cite{JuJan.2014} or full-duplex mode \cite{JuOct.2014}.
The WPCSs also find the applications in device-to-device communications \cite{LiuJan.2016},
secure communications \cite{LiuJan.2016, ShafieOct.2017}, wireless relaying communications \cite{NasirJul.2013, GuNov.2015, DongApr.2016} and passive communications \cite{YaoJan.2018}.
These works mainly focused on the usage of harvested energy to optimize certain performance metrics, such as throughput \cite{JuJan.2014, JuOct.2014}, system secrecy \cite{LiuJan.2016, ShafieOct.2017}, outage probability \cite{NasirJul.2013, GuNov.2015, DongApr.2016}, power consumption \cite{YaoJan.2018}, and energy efficiency (EE) \cite{WuDec.2015, WuMarc2016}.
Another branch of research on RF-EH technology is named as simultaneous wireless information and power transferring (SWIPT) systems, where the energy and information signals are broadcast simultaneously \cite{Varshneyjuly2008, ZhouNov.2013, ClerckxFeb.2018, HuangNov.}.
The concept of SWIPT systems was originally proposed by Varshney in \cite{Varshneyjuly2008}, where he studied the fundamental tradeoff within the \mbox{rate-energy} region in point-to-point (PtP) systems.
Then, the authors in \cite{ZhouNov.2013} proposed two practical receiver architectures, where the RF signals are split into two streams for energy harvesting module and information detecting module.
Moreover, the authors in \cite{ZhouNov.2013, ClerckxFeb.2018} investigated the \mbox{rate-energy} region of PtP systems for linear energy harvester and non-linear energy harvester, respectively.
Since the application of multiple antenna technology can enable the transmission of practical amount of energy, the research on the WPCSs and SWIPT systems was extended to multiple-input-single-output (MISO) \cite{ShiFeb.2016, ShiJun.2014, SonNov.2014, DongApr.2016, BoshkovskaDec.2015} and multiple-input-multiple-output (MIMO) systems \cite{LeeNov.2016, BoshkovskaMay2017, ZhaoFebr2015, XiongAug.2017, RubioJan.2017}.

\subsection{Related Works and Contributions}
The current research on SWIPT systems focus on three types of receivers: 1) time-switching receiver \cite{DongApr.2016, XiongAug.2017}; 2) power-splitting receiver \cite{ShiFeb.2016, ShiJun.2014, XiongAug.2017}; and 3) antenna-separating receiver \cite{SonNov.2014, BoshkovskaDec.2015, ZhaoFebr2015, XiongAug.2017, RubioJan.2017}.
Based on the receiver setups, significant research efforts have been made to investigate the system power minimization  \cite{ShiJun.2014}, system throughput maximization \cite{XiongAug.2017, RubioJan.2017} and harvested energy maximization (HEM) \cite{SonNov.2014, BoshkovskaDec.2015, XiongAug.2017, RubioJan.2017}.
For examples, the authors in \cite{SonNov.2014} and \cite{BoshkovskaDec.2015} studied the HEM for the antenna-separating receiver with linear energy harvester and non-linear energy harvester, respectively.

Though the broadcast characteristic of wireless channels enables one-to-many energy delivery, it also increases the probability of legitimate messages to be eavesdropped in the SWIPT systems \cite{LiuApr.2014, ShiMay2015, NgAug.2014, NgMay2016, Lutobepublished}.
Therefore, the combination of physical layer security with SWIPT systems becomes an emerging research topic.
For examples, the authors in \cite{LiuApr.2014} investigated the secrecy rate maximization (SRM) and HEM in the PtP systems with multiple energy-harvesting nodes (EHNs) co-located with eavesdroppers (EVEs).
Combining Charnes-Cooper transformation \cite{Charnes1962} with one-dimensional search method, the authors in \cite{LiuApr.2014} proposed optimal solutions to the SRM and HEM problems.
Then, the authors in \cite{ShiMay2015} studied the SRM in single-stream and multiple-stream MIMO-PtP systems with perfect channel state information (CSI).
Using a semidefinite relaxation (SDR) technique  and an inexact block-coordinate-descent method, the optimal and near-optimal solutions are obtained for the single-stream and multiple-stream cases.
The authors in \cite{NgSept.2015} studied the secure communication in the renewable resource powered distributed antenna systems with SWIPT capability, which allows each access point to exchange energy with the central processor.
By jointly designing the beamforming vector, artificial noise (AN) covariance matrix and energy exchange variables, the system power consumption is minimized in \cite{NgSept.2015}.

{\color{black}
Due to the tremendous growth of energy consumption, the wireless operators require new approaches to reduce their energy bills or maximize the energy utilization efficiency.
Hence, the EE optimization is one of the most important research issues for the future generation wireless communications \cite{WuAug.2017, LiuDec.2017}.
When multiple EVEs exist in the wireless communication systems, a natural extension to the traditional EE optimization becomes secrecy energy efficiency (SEE) optimization.
The SEE optimization is more important than traditional EE optimization since the traditional EE optimization does not jointly consider the constructive and detrimental effects of artificial noise in energy efficiency.
Although there a number of works considered  EE optimization in SWIPT systems \cite{ShiFeb.2016, ShengApr.2016}, the SEE optimization in SWIPT systems has become an active research area recently \cite{MeiAug.2017}.
Moreover, the SEE optimization induces a better energy utilization efficiency compared with the system throughput maximization or system power minimization.
With perfect CSI, the authors in \cite{MeiAug.2017} investigated the SEE optimization problem in a MIMO-PtP system with multiple EVEs.
Using the Dinkelbach method, they proposed an iterative algorithm to obtain the optimal SEE.
However, their proposed algorithm cannot be applied to the SWIPT systems with multiple legitimate users (LUEs).
When multiple LUEs exist in the MISO-SWIPT systems, the fairness issue among the LUEs becomes an important topic, which has yet been investigated.
}

Since the perfect CSI is challenging to obtain in MISO and MIMO systems\footnote{This situation is especially true when the users are passive or roaming to a new place and do not interchange information with the local BST.}, the authors in \cite{NgAug.2014} studied the system power minimization in the multiuser multiple-input-single-output-multiple-eavesdropper (MISOME)-SWIPT systems with imperfect CSI.
Specifically, they leveraged the energy signals and AN to secure the legitimate communication and satisfy the energy requirement.
Then, the authors extended the work in \cite{NgAug.2014} and jointly considered the transmission power, RF-EH efficiency and interference power leakage as a multiobjective optimization problem with imperfect CSI \cite{NgMay2016}.
In \cite{Lutobepublished}, the authors studied the impact of non-linear energy harvester on system power consumption in a multiuser \mbox{MISOME-SWIPT} system.
To the authors' best knowledge, the robust SEE optimization problem in the multiuser MISOME-SWIPT systems with proportional-secrecy-rate (PSR) constraints has not been reported.

In this work, we consider a multiuser \mbox{MISOME-SWIPT} system where a BST transmits information signals and jamming signals to the LUEs and EVEs, respectively.
Meanwhile, the EHNs scavenge energy from the information signals and jamming signals.
Different from \cite{MeiAug.2017}, we study the SEE optimization via joint design of beamforming vector and AN covariance matrix under the imperfect CSI of EVEs and EHNs.
Moreover, we include the RF-EH constraints and PSR constraints to guarantee the harvested energy at the EHNs and ensure the fairness among LUEs as in \cite{ShenNov.2005, BansalMar.2013}, respectively.
The formulated SEE optimization problem is more complicated than the traditional EE optimization problem \cite{ShiFeb.2016, ShengApr.2016} due to the non-convex PSR constraints and non-convex SEE function.
In order to develop tractable algorithms, we first simplify the SEE optimization problem via exploiting the problem structure.
Based on the introduced parameter, we leverage the SDR technique coupled with a one-dimensional search method to address the simplified SEE optimization problem.
The major contributions of this work can be summarized as follows.
\begin{itemize}
	\item We formulate the SEE maximization problem for a multiuser MISOME-SWIPT system via the joint design of beamforming vector and AN covariance matrix. In order to guarantee the fairness among LUEs, we explicitly confine the secrecy rates of LUEs to a predefined ratio such that the secrecy rates of LUEs are proportional.
        Since the EVEs and EHNs are either roaming users or passive devices, they may not exchange CSI as frequently as that of LUEs.
        Hence, we consider the bounded-error CSI of EVEs and EHNs.
	\item The formulated SEE maximization problem in the MISOME-SWIPT system is non-convex in SEE function and PSR constraints. Therefore, it is difficult and ineffective to solve the non-convex SEE maximization problem via standard methods.
        In order to develop effective algorithms, we exploit the structure of non-convex PSR constraints such that a tractable form of SEE maximization problem is obtained where the strong duality holds. We theoretically prove that: 1) the non-convex PSR constraints can be relaxed into a finite set of convex constraints; and 2) the relaxations are tight.
        Based on these facts, we propose a two-stage algorithm such that an optimal solution is obtained via solving a series of convex optimization problems and a one-dimensional search method.
	\item In order to trade the performance for computational complexity, we propose two suboptimal algorithms based on several heuristic beamforming techniques, e.g., maximal ratio transmission (MRT) and zero-forcing beamforming (ZFBF).
The computational complexity of the proposed optimal and suboptimal algorithms are quantitatively compared.
\end{itemize}

Numerical results are used to verify the performance of the proposed optimal and suboptimal algorithms.

\subsection{Organization and Notations}
The remainder of this paper is organized as follows.
The system model and the SEE maximization problem are presented with imperfect CSI of EVEs and EHNs in Section II.
The optimal and suboptimal algorithms are proposed in Section III and Section IV, respectively.
Numerical results are used to verify the effectiveness of proposed solutions and provide several practical insights in Section V.
Finally, Section VI concludes the work.

\emph{Notations:}
Vectors and matrices are shown in bold lowercase letters and bold uppercase letters, respectively.
$\mathbb{C}^{N\times M}$ denotes the $N\times M$ dimension complex-value matrices.
$\left\|\cdot\right\|_{\F}$ and $\left|\cdot\right|$ are the Frobenius norm  and absolute value, respectively.
$\sim$ stands for ``distributed as''.
$\bm I_{N}$ and $\bm 0_{N\times M}$ denote, respectively, an $N$ dimensional identity matrix and a zero matrix with $N$ rows and $M$ columns.
The expectation of a random variable is denoted as $\mathds{E}\sq{\cdot}$.
$\vec \sq{\bm W}$ obtains a vector by stacking the columns of $\bm W$ under the other.
$\diag\{\bm W_1, \bm W_2, \ldots, \bm W_N\}$ denotes a diagonal matrix with the diagonal elements given by the matrices $\bm W_1, \bm W_2, \ldots, \bm W_N$.
$\left\{\bm{w}_n\right\}_{n \in \mathcal N}$ represents the set made of $\bm{w}_n$, $n \in \mathcal N$.
For a square matrix $\bm W$, $\bm W^{\H}$ and $\mbox{Tr}\br{\bm W}$ denote its conjugate transpose and trace, respectively.
$\bm W \succeq \bm 0$ and $\bm W \succ \bm 0$ respectively denote that $\bm W$ is a positive semidefinite and $\bm W$ is a positive definite matrix.

\section{System Model and Problem Formulation}
\subsection{Overall Description}
\begin{figure}[htb]
\vspace{-0.2 cm}
\centering
  \includegraphics[width= 2.8 in]{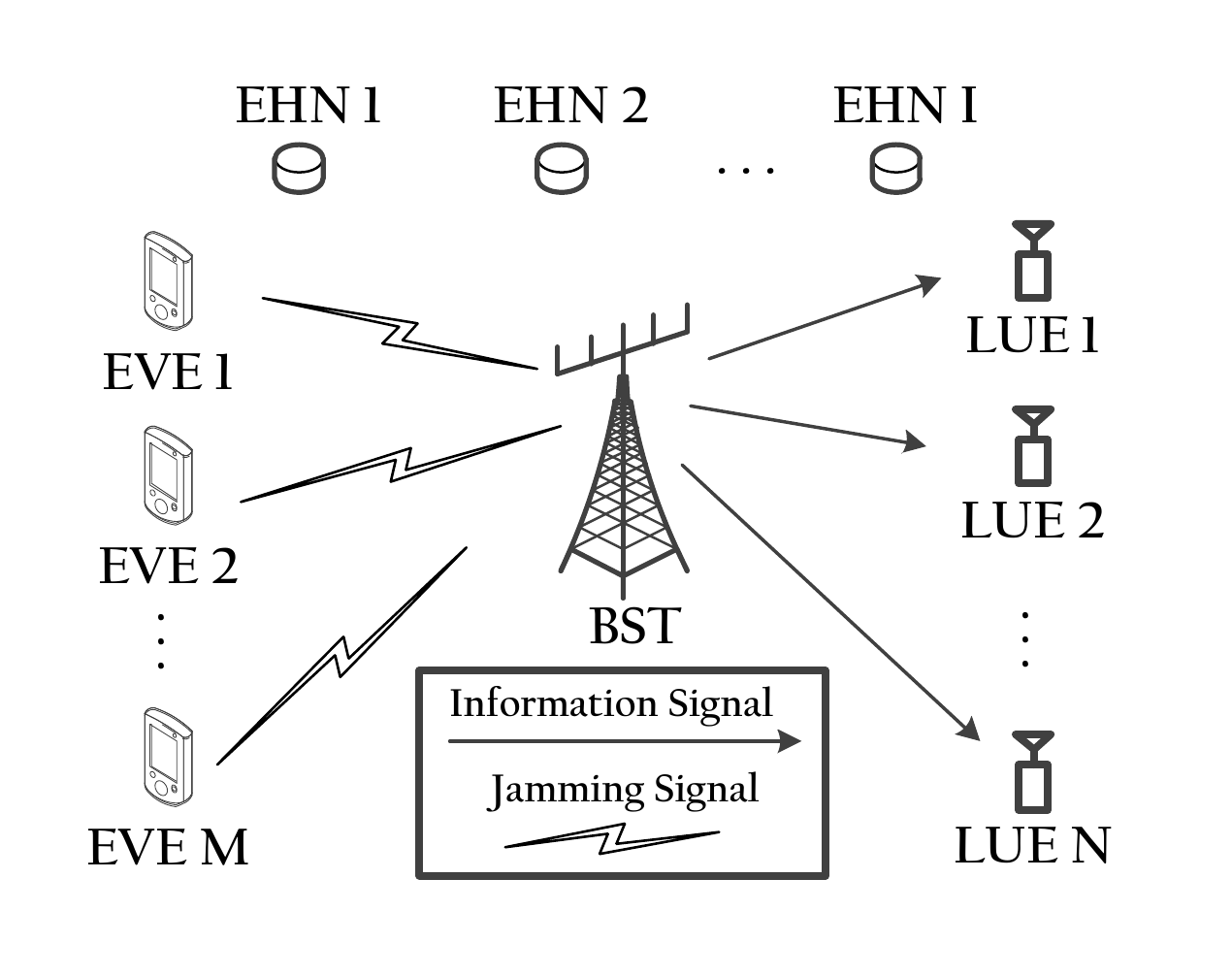}
  \vspace{-0.2 cm}
  \caption{An illustration of the MISOME-SWIPT system that consists one BST, a set of LUEs, a set of EVEs and a set of EHNs.}\label{fg:01}
  \vspace{-0.2 cm}
\end{figure}

We consider downlink transmission of a \mbox{MISOME-SWIPT} system, which consists of one single BST, a set of LUEs, a set of EVEs and a set of EHNs.
Let $\mathcal N = \left\{1, 2, \ldots, N\right\}$, $\mathcal M = \left\{1, 2, \ldots, M\right\}$ and $\mathcal I = \left\{1, 2, \ldots, I\right\}$ denote the set of LUEs, set of EVEs and set of EHNs, respectively.
The BST is equipped with $N_t$ antennas for simultaneous transmission of information and artificial noise over the same single carrier-frequency band.
Therefore, the LUEs and EVEs decode information via beamforming.
Each LUE is equipped with one  antenna to receive the legitimate information.
Each EVE, which is also equipped with one single antenna, can be roaming user from other wireless communication systems and searching for the services from the local BST\footnote{This is due to the fact that the malicious roaming users have the potential to intercept signals intended for LUEs under the coverage of the same BST.}.
Therefore, we leverage the AN to interrupt EVEs and guarantee RF-EH requirement of EHNs.
Here, the EHNs are equipped with one  antenna and can be the passive sensors to monitor the environment status of the MISOME-SWIPT system.

\subsection{Signal Model}
At the BST, the information-bearing signal for the $n$-th LUE is given as $s_n$ with $\mathds{E} \sq{|s_n|^2} = 1$.~Therefore, the transmission signal for LUEs is denoted by
\begin{equation}
\sum\limits_{n = 1}^N {{\bm w_n}{s_n}}
\end{equation}
where $\bm w_n \in \mathbb{C}^{N_t\times 1}$ is the beamforming vector for the $n$-th LUE.~In order to secure the BST-LUE links, the BST needs to transmit an AN vector $\bm q$ to reduce information leakage to EVEs.~Hence, the transmission signal at the BST is given as
\begin{equation}\label{eq:02}
\bm x = \sum\limits_{n = 1}^N {{\bm w_n}{s_n}}  + \bm q
\end{equation}
where the AN vector $\bm q \in \mathbb{C}^{N_t\times 1}$ is modeled as a complex Gaussian random vector with mean zero and covariance $\bm Q \in \mathbb{C}^{N_t\times N_t}$.
In particular, it is assumed that the information-bearing signals $\left\{s_n\right\}_{n \in \cal N}$ and the AN vector $\bm q$ are statistically independent.

\begin{remark}
In order to guarantee the strongly secure communications, the information-bearing signals contain both main information and auxiliary information.
Here, the main information must be reliably delivered for BST-LUE links, and the auxiliary information is used to increase randomness to mislead the EVEs \cite{PierrotSept.2011}.
\end{remark}

The frame-based frequency non-selective fading channels with unit duration for each frame is considered; therefore, the words ``energy'' and ``power'' can be used interchangeably.~In each frame, the BST broadcasts information-bearing signals and jamming signals over the same single carrier-frequency band.~Considering the Rayleigh fading channels, the channel-coefficient vector for the $n$-th BST-LUE link $\bm h_n \in \mathbb{C}^{N_t\times 1}$ is circularly symmetric complex Gaussian (CSCG) distributed as $\bm h_n \sim {\cal CN} \br{\bm0,  \Omega_{u,n}^{-1}\bm I_{N_t}}$.~Here, $\Omega_{u,n}^{-1}$ captures the joint effect of link distance and carrier frequency for the $n$-th BST-LUE link \cite{Goldsmith2005, 3gpp_pathloss}.~As a result, the received signals of the $n$-th LUE is expressed as
\begin{multline}\label{eq:03}
y_{u, n} = \bm h_n^{\H}\bm w_ns_n \\
+  \underbrace{\sumnk {\bm h_n^{\H}{\bm w_k}{s_k}}}_{\mbox{Interference at the $n$-th LUE}} + \underbrace{\bm h_n^{\H}\bm q}_{\mbox{AN}} + z_{u, n}
\end{multline}
where $z_{u, n} \sim {\cal CN} \br{0, \sigma^2_{u, n}}$ is the additive white Gaussian noise (AWGN) with mean zero and variance $\sigma^2_{u, n}$ of the $n$-th LUE.

Based on the received signals in \eqref{eq:03}, the data rate of the $n$-th LUE is denoted as
\begin{equation}\label{eq:04}
R_{u, n} = {\cal BW}\log\br{1+ \mbox{SINR}_{u, n} }
\end{equation}
where
\begin{equation}\label{eq:05}
\mbox{SINR}_{u, n} = \frac{\Tr\br{\bm H_n\bm W_n}}{\sum\limits_{k=1, k \neq n}^N \Tr\br{\bm H_n \bm W_k}  + \Tr\br{\bm H_n\bm Q} + {\sigma _{u, n}^2}}
\end{equation}
with $\bm H_n \triangleq \bm h_n\bm h_n^{\H}$ and $\bm W_n \triangleq \bm w_n\bm w_n^{\H}$, $n \in \cal N$.
In addition, the rank of each beamforming matrix $\bm W_n$ is upper bounded as $\Rank\br{\bm W_n} \le 1$.
The term ${\cal BW}$ denotes the bandwidth of the system.

Besides, the received signal at the $m$-th EVE is denoted as
\begin{multline}\label{eq:06}
y_{e, m} = {\bm g_{e, m}^{\H}}{\bm w_n}{s_n}  \\
+ \underbrace{\sumnk {{\bm g_{e, m}^{\H}}{\bm w_k}{s_k}}}_{\mbox{Interference of the $n$-th LUE}}
+ \underbrace{{\bm g_{e, m}^{\H}}\bm q}_{\mbox{AN}} + z_{e, m}
\end{multline}
where $z_{e, m} \sim {\cal CN}\br{0, \sigma^2_{e, m}}$ is the AWGN with mean zero and variance $\sigma^2_{e, m}$ at the $m$-th EVE, $m \in \cal M$;~$\bm g_{e, m} \in \mathbb{C}^{N_t \times 1}$ is the channel-coefficient vector for the $m$-th BST-EVE link.

Based on the signal model in \eqref{eq:06}, the information leakage to the $m$-th EVE of the $n$-th LUE by treating the interference as noise is denoted as \cite{Goldsmith2005}
\begin{equation}\label{eq:07}
R_{e, m \rightarrow n} = {\cal BW}\log\br{1 + \SINR_{e, m \rightarrow n}}
\end{equation}
where
\begin{equation}\label{eq:08}
\SINR_{e, m\rightarrow n} = \frac{\Tr\br{\bm G_{e, m}\bm W_n}}{\sum\limits_{k=1, k \neq n}^N \Tr\br{\bm G_{e, m} \bm W_k}  + \Tr\br{\bm G_{e, m}\bm Q} + {\sigma _{e, m}^2}}
\end{equation}
where $\bm G_{e, m} \triangleq \bm g_{e, m}\bm g_{e, m}^{\H}$.

Denote the channel-coefficient vector of the $i$-th BST-EHN link as $\bm g_{h, i} \in \mathbb{C}^{N_t\times 1}$, the received baseband signal of the $i$-th EHN is denoted as
\begin{equation}\label{eq:09}
y_{h, i} = \sumn\bm g_{h, i}^{\H}\bm w_n s_n + \bm g_{h, i}^{\H}\bm q + z_{h, i}
\end{equation}
where $z_{h, i} \sim {\cal CN}\br{0, \sigma_{h,i}^2}$ is the AWGN at the $i$-th EHN, and $\bm g_{h, i}$ is associated with the pathloss and carrier frequency of the $i$-th BST-LUE link.

Ignoring the energy that can be harvested from the AWGN, the amount of harvested energy of the $i$-th EHN is denoted as
\begin{equation}\label{eq:10}
P_{h, i} = \xi_{h, i}\br{\sumn\Tr\br{\bm G_{h, i}\bm W_n} + \Tr\br{\bm G_{h, i}\bm Q}}
\end{equation}
where $\bm G_{h, i} \triangleq  \bm g_{h, i}\bm g_{h, i}^{\H}$, and $\xi_{h, i} \in \left(0, 1\right]$ is the energy conversion efficiency.

\subsection{Secure Data Rate, Power Consumption and SEE}
{\color{black}
Since we consider strongly secure communication, the information-leakage rate of the \mbox{$n$-th} BST-LUE link is confined to be smaller than or equal to the rate of auxiliary information $R_{e, n}^{\REQ}$.~Hence, the secrecy rate of the $n$-th BST-LUE link is written  as \cite{PierrotSept.2011, Dong2018a}
\begin{equation}\label{eq:11}
R_{u, n}^{\SEC} = \br{R_{u, n} - R^{\REQ}_{e, n}}^+
\end{equation}
with
\begin{equation}\label{eq:12}
R_{e, m\rightarrow n} \le R^{\REQ}_{e, n}, \forall m
\end{equation}
where $\br{\cdot}^+ = \max\br{\cdot, 0}$.}

Furthermore, the power consumption of BST in the MISOME-SWIPT system is comprised of the power of downlink beamforming, power of AN and power of circuit.~Therefore, the expression of power consumption is given as
\begin{equation}\label{eq:13}
P^{\TOT} = \frac{1}{\phi }\br{ {\sumn \Tr\br{\bm W_n}  + \Tr \br{\bm Q} } } + {P^{\C}}
\end{equation}
where the constant circuit power consumption $P^{\C}$ is given as $P^{\C} = P^{\mbox{\tiny SP}}\br{0.87 + 0.1N_t + 0.03N_t^2}$ since the circuit power consumption is positively correlated to the number of antennas \cite{AttarOct.2011}.
Here,  the term $P^{\mbox{\tiny SP}}$ denotes the baseband processing power consumption.

In this work, the SEE is defined as the ratio of the sum secrecy rates of BST-LUE links over the power consumption of BST.~Based on \eqref{eq:11} and \eqref{eq:13}, the SEE of \mbox{MISOME-SWIPT} system is given as
\begin{equation}\label{eq:14}
\mbox{SEE}\br{\left\{\bm W_n\right\}_{n \in \cal N}, \bm Q} = \frac{\sumn R_{u, n}^{\SEC}}{{P^{\TOT}}}
\end{equation}
whose unit is Nats/joule.

\begin{remark}
Different from \cite{ShiFeb.2016}, we ignore the harvested energy in the power consumption and investigate the SEE maximization.~This is due to the fact that the harvested energy is stored in the integrated battery of EHNs and cannot reduce the energy bills of the wireless operators.
\end{remark}

{\color{black}
\subsection{Channel State Information}
We assume that the MISOME-SWIPT system operates in time-division duplex mode.~At the beginning of each frame, each LUE sends a pilot signal to the BST.~After receiving the pilot signals, the BST estimates the channels associated with each LUE that sends the pilot signal.~Since each LUE facilitates each uplink transmission with pilot signals, the BST can exploit the uplink reciprocity to periodically update the downlink CSI for BST-LUE links.~Therefore, we assume that the BST can perfectly estimate the downlink CSI for the BST-LUE links.~However, the EVEs and EHNs can be the roaming users and/or low-power nodes and do not exchange the pilot signals with the BST as frequently as the LUEs.~The obtained CSI for the BST-EVE links can be imperfect subject to channel estimation errors and/or quantization errors.~Hence, we are motivated to use the bounded-error CSI model to formulate the CSI for the BST-EVE and BST-EHNs links as \cite{WangAug.2009, SunAug.2016}
\begin{align}
\bm g_{e, m} &= \bar{\bm g}_{e, m} + \Delta{\bm g}_{e, m}, \left\|\Delta\bm g_{e, m} \right\|_{\F} \le \Theta_{e, m}, \forall m \\
\bm g_{h, i} &= \bar{\bm g}_{h, i} + \Delta{\bm g}_{h, i}, \left\|\Delta{\bm g}_{h, i}\right\|_{\F} \le \Theta_{h, i}, \forall i
\end{align}
where $\bar{\bm g}_{e, m} \in \mathbb{C}^{N_t \times 1}$ and $\Delta\bm g_{e, m} \in \mathbb{C}^{N_t \times 1}$ are, respectively, the estimated channel-coefficient vector and channel uncertainty of the $m$-th BST-EVE link.
Similarly, $\bm g_{h,i}$ and $\Delta\bm g_{h,i}$ are the estimated channel-coefficient vector and channel uncertainty vector of the $i$-th EHN.
The positive constants $\Theta_{e, m}$ and $\Theta_{h, i}$ denote the radii of uncertainty region of $\bm g_{e, m}$ and $\bm g_{h, i}$.
Here, the vectors $\bar{\bm g}_{e, m} \sim {\cal CN}\br{\bm0,  \Omega_{e, m}^{-1}\bm I_{N_t}}$ and $\bar{\bm g}_{h, i} \sim {\cal CN}\br{\bm 0,  \Omega_{h, i}^{-1}\bm I_{N_t}}$ are CSCG distributed having the terms $\Omega_{e,m}$ and $\Omega_{h, i}$ that take into account of the joint effect of pathloss and carrier frequency of the $m$-th BST-EVE link.
In addition, the channel uncertainty vectors $\Delta\bm g_{e, m}$ and $\Delta{\bm g}_{h, i}$ capture the joint effect of estimation errors and time-varying characteristics of wireless channels.

\begin{remark}
When the roaming users are malicious and want to intercept the unauthorized services, they become the EVEs.
However, the EVEs also need to exchange pilot signal with the BST to establish links to obtain their registered services. The low-power nodes can be sensors in the coverage of the same BST. Although the low-power nodes do not interact with the BST, they may still need to report messages to the data fusion. Thus, the BST can still estimate the CSI of BST-EHN links.
\end{remark}
}

\subsection{Problem Formulation}
Our objective is to maximize SEE via the joint design of beamforming vector and AN covariance matrix in the MISOME-SWIPT system.~At the beginning of each frame, the BST performs beamforming and jamming in a centralized way.~The SEE maximization problem is formulated as
\begin{subequations}\label{eq:15}
\begin{align}
\max\limits_{\left\{ \bm W_n \right\}_{n \in \cal N},\bm Q} &\; \mbox{SEE}\br{\left\{ {{\bm W_n}} \right\}_{n \in \cal N},\bm Q} \label{eq:15a}\\
\st&\; {R_{u, 1}^{\SEC}}: \ldots :{R_{u, N}^{\SEC}} = {\varphi _1}: \ldots :{\varphi _N}\label{eq:15b}\\
&\; \max\limits_{\left\|\Delta\bm g_{e, m}\right\|_{\F} \le \Theta_{e, m}} R_{e, m\rightarrow n} \le R_{e, n}^{\REQ}, \forall m, n \label{eq:15c}\\
&\; \min\limits_{\left\|\Delta\bm g_{h, i}\right\|_{\F} \le \Theta_{h, i}} P_{h, i} \ge P_{h, i}^{\REQ}, \forall i \label{eq:15d}\\
&\; \sumn \Tr\br{\bm W_n} +\Tr\left(\bm Q\right) \le {P^{\max }}\label{eq:15e}\\
&\; \bm Q \succeq \bm 0, \bm W_n \succeq \bm 0,  \forall n \label{eq:15f} \\
&\; \Rank\br{\bm W_n} \le 1, \forall n \label{eq:15g}
\end{align}
\end{subequations}
where the constraints in \eqref{eq:15b} are used to guarantee the proportional fairness on the secrecy rates of LUEs with $\sum\nolimits_{n=1}^N \varphi_n = 1$ and $\varphi_n \ge 0$ \cite{ShenNov.2005, BansalMar.2013}; and the information-leakage constraints and the harvested-power constraints are respectively defined in \eqref{eq:15c} and \eqref{eq:15d}.~Here, the term $P_{h, i}^{\REQ}$ in \eqref{eq:15d} denotes the energy requirement of the $i$-th EHN; and the constant $P^{\max}$ in the constraints in \eqref{eq:15e} is the power budget of BST due to the circuit limitation.

\begin{remark}
Although we use the linear energy harvester, it is easy to justify that, by properly setting the threshold $P_{h, i}^{\REQ}$, the non-linear energy harvester model can be incorporated in \eqref{eq:15d} \cite{BoshkovskaDec.2015, DongApr.2016}.~Using the model in \cite{BoshkovskaDec.2015}, harvesting $P_{h, i}^{\REQ}$ power requires the input power $\hat P_{h, i}^{\REQ} = \frac{\xi_{h, i}}{a_{h, i}}\log\left(\frac{M_{h, i} + P_{h, i}^{\REQ}\exp\left(a_{h, i}b_{h, i}\right)}{M_{h, i} - P_{h, i}^{\REQ}}\right)$, where $M_{h, i}$, $a_{h, i}$ and $b_{h, i}$ are the shaping parameters \cite{BoshkovskaDec.2015}.
\end{remark}

{\color{black}
\section{Optimal SEE Maximization}
We observe that the finding the optimal solution to optimization problem \eqref{eq:15} is challenging  due to the non-convex objective function \eqref{eq:15a}, the PSR constraints in \eqref{eq:15b}, the infinite amount of information-leakage constraints in \eqref{eq:15c} and RF-EH constraints in \eqref{eq:15d} and the non-convex rank-one constraints in \eqref{eq:15g}.
In order to solve the problem \eqref{eq:15}, we deal with non-convexity in the objective function \eqref{eq:15a} and constraints in \eqref{eq:15b} and \eqref{eq:15g} via a convex relaxation and SDR.
Moreover, we obtain the equivalent convex forms in finite amount for the constraints in \eqref{eq:15c} and \eqref{eq:15d} via $\cal S$-procedure \cite{boyd2004convex}.

Since the non-convex objective function \eqref{eq:15a} and PSR constraints in \eqref{eq:15b} contain the secrecy rate of LUEs, we are motivated to handle the non-convexity in \eqref{eq:15a} and \eqref{eq:15b} via convex relaxation.
Introducing an auxiliary variable $t$ with $0 \le t \le \sum\nolimits_{n=1}^N R_{u,n}^{\SEC}$, the objective function \eqref{eq:15a} and PSR constraints \eqref{eq:15b} are respectively relaxed as
\begin{equation}\label{eq:19}
\mbox{SEE}\br{\left\{\bm W_n\right\}_{n \in \cal N}, \bm Q} \ge \frac{t}{P^{\TOT}\br{\left\{\bm W_n\right\}_{n \in \cal N}, \bm Q}}
\end{equation}
and
\begin{multline}\label{eq:20}
\frac{1 + \theta_n\br{t}}{\theta_n\br{t}}\Tr\br{\bm H_n \bm W_n} \ge \sumk \Tr\br{\bm H_n \bm W_k} \\
+ \Tr\br{\bm H_n \bm Q} + \sigma_{u, n}^2, \forall n
\end{multline}
where ${\theta _n}\br{t} \triangleq \exp \br{\frac{\varphi_n}{\cal BW}  t + \widetilde{R}^{\REQ}_{e, n}} - 1$.
Note that the sum secrecy rate of all BST-LUE links is lower bounded by $t$, and the secrecy rate of the $n$-th BST-LUE link is lower bounded by $\varphi_n t$, $n \in \cal N$.
Thus, the lower bounds of secrecy rates of BST-LUE links satisfy the predefined ratios.

Therefore, the convex-relaxation version of optimization problem \eqref{eq:15} is obtained as
\begin{subequations}\label{eq:21}
\begin{align}
\max \limits_{{\left\{ {\bm W_n} \right\}_{n \in {\cal N}}}, \bm Q, t} &\;
\frac{t}{P^{\TOT}\br{\left\{\bm W_n\right\}_{n \in \cal N}, \bm Q}} \label{eq:21a}\\
\st&\; \eqref{eq:15c}-\eqref{eq:15g} \mbox{ and } \eqref{eq:20}. \label{eq:21b}
\end{align}
\end{subequations}

In problem \eqref{eq:21}, we replace the objective function \eqref{eq:15a} with its lower bound \eqref{eq:19} and relax the PSR constraints in \eqref{eq:15c} by \eqref{eq:20}.
Before we proceed to justify the equivalence between \eqref{eq:15} and \eqref{eq:21}, we verify the tightness of convex relaxations in \eqref{eq:19} and \eqref{eq:20} for problem \eqref{eq:21} in Proposition \ref{pr:01}.

\begin{proposition}\label{pr:01}
Suppose that the right-hand-side of \eqref{eq:19} and the constraints in \eqref{eq:20} are not inactive under the optimal beamforming matrices $\{\bm W_n^*\}_{n \in \cal N}$ and AN covariance matrix $\bm Q^*$ to problem \eqref{eq:21}.
Then, the inequalities in \eqref{eq:19} and \eqref{eq:20} are active via another optimal beamforming matrices $\{\widehat{\bm W}_n^*\}_{n \in \cal N}$ and AN covariance matrix $\widehat{\bm Q}^*$ to \eqref{eq:21}, which can be constructed as
\begin{subequations}\label{eq:pr1:01}
\begin{align}
\widehat{\bm W}_n^* &= y_n \bm W_n^*, \forall n \label{eq:pr1:01a}\\
\widehat{\bm Q}^* &= \bm Q^* + \sumn \br{1-y_n} \bm W_n^*  \label{eq:pr1:01b}\\
y_n &= \frac{\theta_n\br{t^*}}{1 + \theta_n\br{t^*}}\frac{\Tr\br{\bm H_n\br{\sumn \bm W_n^* + \bm Q^*}}  + \sigma_{u,n}^2}{\Tr
\br{\bm H_n \bm W_n^*}}. \label{eq:pr1:01c}
\end{align}
\end{subequations}
\end{proposition}
\begin{IEEEproof}
See Appendix \ref{apdx:01}.
\end{IEEEproof}

Based on Proposition \ref{pr:01}, we conclude the equivalence between the problems \eqref{eq:15} and \eqref{eq:21}, which is verified in Appendix \ref{apdx:01a}.
Hence, we investigate the problem \eqref{eq:21} to seek a  low complexity algorithm for the problem \eqref{eq:15}.

In order to deal with the infinite amount of information-leakage constraints in \eqref{eq:15c} and \eqref{eq:15d}, we first review the $\cal S$-procedure \cite{boyd2004convex} in Lemma \ref{le:01}.

\begin{lemma}[$\cal S$-procedure \cite{boyd2004convex}]\label{le:01}
Denote the functions $f_1\br{\bm x}$ and $f_2\br{\bm x}$ as $f_1\br{\bm x} = \bm x^{\H}\bm A_1\bm x + \bm b_1^{\H}\bm x + \bm x^{\H}\bm b_1 + c_1$ and $f_2\br{\bm x} = \bm x^{\H}\bm A_2\bm x + \bm b_2^{\H}\bm x + \bm x^{\H}\bm b_2 + c_2$ with $\bm A_1 = \bm A_1^{\H}$ and $\bm A_2 = \bm A_2^{\H}$.
Then, the condition $f_1\br{\bm x} \le 0 \Rightarrow f_2\br{\bm x} \le 0$ holds if and only if there is $\eta \ge 0$ such that
\begin{equation}\label{eq:le1:01}
\eta\sq{ {\begin{array}{*{20}{c}}
{\bm A_1}&{\bm b_1}\\
{\bm b_1^{\H}}&{c_1}
\end{array}} } -
\sq{{\begin{array}{*{20}{c}}
{\bm A_2}&{\bm b_2}\\
{\bm b_2^{\H}}&{c_2}
\end{array}}}  \succeq \bm 0
\end{equation}
provided that there exists a point $\bm x$ such that $f_k\br{\bm x} < 0$, $k = 1, 2$.
\end{lemma}

With \eqref{eq:07} and \eqref{eq:08}, we obtain an equivalent form of information-leakage constraints in \eqref{eq:15c} as
\begin{equation}\label{eq:16}
\bm g_{e, m}^{\H} \bm X_n \bm g_{e, m} \le \sigma_{e, m}^2, \forall \Delta\bm g_{e, m}, \forall m, n
\end{equation}
where $\bm X_n \triangleq \frac{1}{1 - \exp\br{-\widetilde{R}^{\REQ}_{e, n}}}\bm W_n - \sum\nolimits_{k=1}^N \bm W_k - \bm Q$ with $\widetilde{R}^{\REQ}_{e, n} =  \frac{R_{e, m\rightarrow n}^{\REQ}}{\cal BW}$ and $\left\|\Delta \bm g_{e, m}\right\|_{\F}^2 \le \Theta^2_{e, m}$.

Applying Lemma \ref{le:01} to \eqref{eq:16}, we obtain an equivalent form of information-leakage constraints in \eqref{eq:15c} as
\begin{multline}\label{eq:17}
\bm\Phi_{m,n}\br{\bm X_n, \zeta_{m, n}} = \sq{ {\begin{array}{*{20}{c}}
\zeta_{m, n}\bm I_{N_t}& \bm 0 \\
\bm 0 & \sigma_{e,m}^2 - \zeta_{m, n}\Theta_{e,m}^2
\end{array}} } \\
- \widetilde{\bm G}_{e, m}^{\H} \bm X_n \widetilde{\bm G}_{e, m} \succeq \bm 0, \forall m, n
\end{multline}
where $\widetilde{\bm G}_{e, m} \triangleq \sq{\bm I_{N_t}, \bar{\bm g}_{e,m}}$, and $\zeta_{m, n} \ge 0$ is the introduced auxiliary variable, $m \in \cal M$ and $n \in \cal N$.

Similarly, the equivalent form of \eqref{eq:15d} is obtained as
\begin{equation}\label{eq:18temp}
\bm g_{h,i}^{\H} \bm Y \bm g_{h,i} \ge \frac{P_{h,i}^{\REQ}}{\xi_{h,i}}, \forall i
\end{equation}
with $\bm Y = \sum\nolimits_{n=1}^N\bm W_n + \bm Q$.

Then, applying Lemma 1 to \eqref{eq:18temp} and performing some algebraic manipulations, we obtain an equivalent form of RF-EH constraints in \eqref{eq:15d} as
\begin{multline}\label{eq:18}
\bm \Psi_i\br{\bm Y, \eta_i} = \sq{ {\begin{array}{*{20}{c}}
\eta_i\bm I_{N_t}& \bm 0 \\
\bm 0 & -\frac{P_{h,i}^{\REQ}}{\xi_{h,i}} - \eta_i\Theta_{h,i}^2
\end{array}} } \\
+ \widetilde{\bm G}_{h, i}^{\H} \bm Y \widetilde{\bm G}_{h, i} \succeq \bm 0, \forall i
\end{multline}
where $\widetilde{\bm G}_{h, i} = \sq{\bm I_{N_t}, \bm g_{h,i}}$, and $\eta_i$ is the introduced nonnegative auxiliary variable, $i \in \cal I$.

Based on the previous manipulations in \eqref{eq:20}, \eqref{eq:17} and \eqref{eq:18}, we can now convexify the feasible region of the problem \eqref{eq:21} via SDR technique.
By dropping the constraints in \eqref{eq:15g}, the feasible region of problem \eqref{eq:21} is transformed into a convex one as
\begin{subequations}\label{eq:24}
\begin{align}
\max \limits_{\scriptstyle{\left\{ {\bm W_n} \right\}_{n \in {\cal N}}}, \bm Q, \{\eta_{i}\}_{i \in \cal I}\hfill\atop
\scriptstyle{\left\{ {{\zeta _{m,n}}} \right\}_{m \in {\cal M},n \in {\cal N}}, t}\hfill} &\;
\frac{t}{P^{\TOT} \br{\left\{\bm W_n\right\}_{n \in \cal N}, \bm Q}} \label{eq:24a}\\
\st&\; \zeta_{m, n} \ge 0, \eta_i \ge 0, \forall m, n, i \label{eq:24b}\\
&\; \eqref{eq:15e}, \eqref{eq:15f}, \eqref{eq:20}, \eqref{eq:17} \mbox{ and } \eqref{eq:18}. \label{eq:24d}
\end{align}
\end{subequations}

\begin{remark}
The only challenge in solving the problem \eqref{eq:24} is to deal with the non-convex objective function.
Since the objective function \eqref{eq:24a} is joint convex with respect to the beamforming matrices $\{\bm W_n\}_{n \in \cal N}$, AN covariance matrix $\bm Q$ and auxiliary variables $\{y_i\}_{i \in \cal I}$ and $\left\{ {{\zeta _{m,n}}} \right\}_{m \in {\cal M},n \in {\cal N}}$, we propose a two-stage algorithm to solve the problem \eqref{eq:24}: 1) solving the problem \eqref{eq:24} with given $t$; 2) performing a one-dimensional search to obtain the optimal $t^*$.
\end{remark}

\begin{remark}
The feasible region of problem \eqref{eq:24} is a convex hull of the feasible region of problem \eqref{eq:21}.
Hence, the optimal value of \eqref{eq:24} serves as the lower bound of the optimal value of \eqref{eq:21} with the optimal $t^*$.
Thus, we are motivated to show that the performed SDR in \eqref{eq:24} given $t^*$ is tight such that the optimal solution to \eqref{eq:24} given $t^*$ is the optimal value of \eqref{eq:21}.
\end{remark}

%

Due to the power budget constraint in \eqref{eq:15e}, the strong duality of problem \eqref{eq:24} may not hold.
Hence, we are motivated to figure out an equivalent form of problem \eqref{eq:24}, whose strong duality holds.
We observe that the power constraint in \eqref{eq:15e} and objective function \eqref{eq:24a} have a common term $\sum\nolimits_{n=1}^N\Tr\br{\bm W_n} + \Tr\br{\bm Q}$.
Dropping the power constraint in \eqref{eq:15e} and fixing the value of $t$, we solve the problem \eqref{eq:24} as
\begin{subequations}\label{eq:26}
\begin{align}
\min\limits_{\scriptstyle{\left\{ {\bm W_n} \right\}_{n \in {\cal N}}}, \bm Q, \{\eta_{i}\}_{i \in \cal I}\hfill\atop
\scriptstyle{\left\{ {{\zeta _{m,n}}} \right\}_{m \in {\cal M},n \in {\cal N}}}\hfill} &\;
\sumn \Tr\br{\bm W_n} + \Tr\br{\bm Q} \label{eq:26a}\\
\st&\; \zeta_{m, n} \ge 0, \eta_i \ge 0, \forall m, n, i \label{eq:26b}\\
&\; \eqref{eq:15f}, \eqref{eq:20}, \eqref{eq:17} \mbox{ and } \eqref{eq:18}. \label{eq:26d}
\end{align}
\end{subequations}

The benefits of dropping the power constraint in \eqref{eq:15e} are three folds: 1) the problem \eqref{eq:26} is always feasible; 2) the strong duality of \eqref{eq:26} always holds \cite{boyd2004convex}; and 3) the computational complexity of solving \eqref{eq:26} is lower than that of \eqref{eq:24}.
Since the problem \eqref{eq:24} is feasible if and only if the optimal value of \eqref{eq:26} is less than or equal to $P^{\max}$, we exploit the monotonicity of objective function \eqref{eq:26a} with respect to $t$ via the Proposition \ref{pr:02}.}

\begin{proposition}\label{pr:02}
Let $f\br{t} = \sum\nolimits_{n=1}^N\Tr\br{\bm W_n^t} + \Tr\br{\bm Q^t}$ denote the optimal power consumption of \eqref{eq:26} given $t$ where $\left\{\bm W_n^t\right\}_{n \in \cal N}$ and $\bm Q^t$ are, respectively, the optimal beamforming matrices and AN covariance matrix to \eqref{eq:26} given $t$.
The optimal power consumption $f\br{t}$ monotonically increases  with $t$.
\end{proposition}
\begin{IEEEproof}
See Appendix \ref{apdx:02}.
\end{IEEEproof}

With the Proposition \ref{pr:02}, we can claim that there exists only one value of $t^{\max}$ such that the optimal power consumption $f\br{t^{\max}} = P^{\max}$.
The value of $t^{\max}$ can be obtained  via a one-dimensional line search.
Now, we investigate the optimality of SDR technique.
Hereinafter, we ignore the superscript $t$ in order to keep the brevity of presentation.

\begin{proposition}\label{pr:03}
There exists an optimal solution $\br{\{\bm W_n\}_{n \in \cal N}, \bm Q, \{\eta_i\}_{i \in \cal I}, \{\zeta_{m, n}\}_{{m\in\cal M}, {n\in\cal N}}}$ to the problem \eqref{eq:26} with the rank of beamforming matrices $\{\bm W_n\}_{n\in \cal N}$ satisfying
\begin{equation}\label{eq:pr:01}
\Rank\br{\bm W_n} \le 1, \forall n.
\end{equation}
\end{proposition}
\begin{IEEEproof}
See Appendix \ref{apdx:03}.
\end{IEEEproof}

When the value of $t$ guarantees $f\br{t} \le P^{\max}$, the optimal solution to problem \eqref{eq:24} is obtained via solving the problem \eqref{eq:26} and performing the one-dimensional search for the optimal $t^*$ based on Proposition \ref{pr:03}.
As a result, we propose \textbf{Algorithm} \ref{alg:01}, which is named as SDP empowered two-stage beamforming and artificial jamming (SDP-TsBAJ) algorithm.

\begin{algorithm}[ht]\small
  \centering
  \caption{SDP Empowered Two-Stage Beamforming and Artificial Jamming Algorithm}\label{alg:01}
  \begin{algorithmic}[1]
  \State BST initializes $t = 0$, $\Delta t$ and $\mbox{SEE}\br{\Delta t} = 0$ with unit Nats/joule
  \Repeat
  \State BST obtains ${\OPT}\br{t}$ via solving \eqref{eq:26}
  \State BST generates the augmented SEE sequence as $\mbox{ASEE}\br{t} = \max\{\mbox{SEE}\br{t-\Delta t}, \frac{t}{P^{\TOT}}\}$ to store the optimal SEE  given $t$
  \State $t \leftarrow t + \Delta t$
  \Until{$\sum\nolimits_{n=1}^N\Tr\br{\W_n} + \Tr\br{\Q} > P^{\max}$ and/or the sequence $\mbox{ASEE}\br{t}$ converges}
  \If{The constraints in \eqref{eq:20} are inactive}
  \State BST performs the feasibility recovery procedure in \eqref{eq:pr1:01a}--\eqref{eq:pr1:01c}
  \EndIf
  \If{$\Rank\br{\bm W_n^*} > 1$}
  \State BST performs the rank-one recovery procedure in \eqref{eq:apdx03:10a} and \eqref{eq:apdx03:10b}
  \EndIf
  \end{algorithmic}
\end{algorithm}

We observe that the generated augmented SEE sequence increases monotonically with the number of search points.
Since the system SEE is upper bounded due to the maximum transmit power of BST, we conclude that the SDP-TsBAJ algorithm in  converges.

\section{Low Complexity Suboptimal SEE Maximization}
Although the proposed SDP-TsBAJ algorithm can significantly reduce the computational complexity compared with the brand-and-bound algorithm, we observe that the computational complexity of SDP-TsBAJ algorithm increases fast with the number of LUEs.
In order to reduce the computational complexity of the SDP-TsBAJ algorithm, we design the beamforming vector based on heuristic beamforming techniques: ZFBF and MRT.
Denote the matrices $\bm H$ and $\bm H_{-n}$ as $\bm H = \sq{\bm h_1, \bm h_2, \ldots, \bm h_N}$ and $\bm H_{-n} = \sq{\bm h_1, \ldots, \bm h_{n-1}, \bm h_{n+1}, \ldots, \bm h_N}$, respectively.
To remove the interference of AN, we perform the singular value decomposition to the matrix $\bm H^{\H}$ as $\bm \Phi \in \mathbb{C}^{N_t\times \br{N_t - N}}$ with $\bm \Phi^{\H}\bm \Phi = \bm I_{N_t - N}$ and construct the AN covariance matrix as
\begin{equation}\label{eq:sb:01}
\bm Q = \bm \Phi \overline{\bm Q} \bm \Phi^{\H}
\end{equation}
where the matrix $\overline{\bm Q} \in \mathbb{C}^{(N_t - N) \times (N_t - N)}$.

Similarly, we apply the ZFBF technique to the beamforming vectors of LUEs as
\begin{equation}\label{eq:sb:02}
\bm w_n = \bm \Xi_n\overline{\bm w}_n, \forall n
\end{equation}
where the matrix $\bm \Xi_n \in \mathbb{C}^{N_t \times(N_t - N + 1)}$ is obtained via singular value decomposition of the matrix $\bm H_{-n}^{\H}$ with $\bm \Xi_n^{\H}\bm \Xi_n = \bm I_{N_t - N + 1}$.
Here, each entry of $\overline{\bm w}_n \in \mathbb{C}^{\br{N_t - N + 1}\times 1}$ is a complex weight to each column vector of $\bm\Xi_n$.

Substituting \eqref{eq:sb:01} and \eqref{eq:sb:02} into the problem \eqref{eq:26}, we obtain a simplified problem as
\begin{subequations}\label{eq:sb:03}
\begin{align}
\mathop {\min }\limits_{\scriptstyle{\{\overline{\bm W}_n\}_{n \in {\cal N}}, \overline{\bm Q}, \{\eta_i\}_{i \in \cal I}}\hfill\atop
\scriptstyle{\left\{ {{\zeta _{m,n}}} \right\}_{m \in {\cal M},n \in {\cal N}}}\hfill} &\;
\sumn \Tr\br{\overline{\bm W}_n} + \Tr\br{\overline{\bm Q}} \label{eq:sb:03a}\\
\st&\; \zeta_{m, n} \ge 0, \overline{\bm W}_n\succeq \bm 0, \forall m, n \label{eq:sb:03b}\\
&\; \eta_i \ge 0, \overline{\bm Q}  \succeq \bm 0, \forall i\\
&\; \bm\Phi_{m,n}\br{\overline{\bm X}_n, \zeta_{m,n}} \succeq \bm 0, \forall m, n\\
&\; \bm \Psi_i\br{\overline{\bm Y}, \eta_i} \succeq \bm 0, \forall i \\
&\; \Tr\br{\bm\Xi_n^{\H}\bm H_n\bm\Xi_n \overline{\bm W}_n} \nonumber\\
&\; \hspace{1.8 cm} = \theta_n\br{t}\sigma_{u, n}^2, \forall n  \label{eq:sb:03c}
\end{align}
\end{subequations}
where the matrices $\overline{\bm X}_n$ and $\overline{\bm Y}$ are, respectively, denoted as
\begin{multline}\label{eq:sb:04}
\overline{\bm X}_n = \frac{1}{1 - \exp\br{-\widetilde{R}^{\REQ}_{e, m\rightarrow n}}}\bm\Xi_n \overline{\bm W}_n\bm\Xi_n^{\H} \\
- \sum\limits_{k=1}^N \bm\Xi_k \overline{\bm W}_k\bm\Xi_k^{\H} - \bm\Phi\overline{\bm Q}\bm\Phi^{\H}, \forall n
\end{multline}
and
\begin{equation}\label{eq:sb:05}
\overline{\bm Y} = \sum\limits_{n=1}^N \bm\Xi_n \overline{\bm W}_n\bm\Xi_n^{\H} + \bm\Phi\overline{\bm Q}\bm\Phi^{\H}
\end{equation}
with $\overline{\bm W}_n = \overline{\bm w}_n\overline{\bm w}_n^{\H}$.

The optimization problem \eqref{eq:sb:03} is convex and solvable in CVX.
The rank-one constraints for the matrices $\overline{\bm W}_n$ can be recovered following the procedure in \cite{Dong2018a}.
Instead of solving \eqref{eq:26}, the proposed ZFBF-TsBAJ algorithm solves \eqref{eq:sb:03} at the third step of the SDP-TsBAJ algorithm.

In order to obtain the beamforming vector $\bm w_n$ in ZFBF-TsBAJ algorithm, we still need to calculate a complex weight vector $\overline{\bm w}_n$.
Thus, we leverage the MRT technique to reduce number of variables such that computational complexity of ZFBF-TsBAJ algorithm is further reduced.
Performing the MRT technique to the $n$-th ZFBF vector by setting $\bm w_n = \bm h_n$, we obtain $\overline{\bm w}_n = \bm\Xi_{n}^{\H}\bm h_n$.
Hence, the $n$-th MRT-ZFBF beamforming vector is given as
\begin{equation}\label{eq:sb:06}
\bm w_n = \sqrt{p_n}\frac{\bm \Xi_n\bm\Xi_{n}^{\H}\bm h_n}{\left\|\bm\Xi_{n}^{\H}\bm h_n\right\|_{\F}}, \forall n
\end{equation}
where $p_n$ is the transmit power for the $n$-th BST-LUE link.

Substituting \eqref{eq:sb:01} and \eqref{eq:sb:06} into \eqref{eq:20}, we obtain the closed-form transmit power for the $n$-th BST-LUE link as
\begin{equation}
p_n = \frac{\theta_n\br{t}\sigma_{u,n}^2}{\left\|\bm\Xi_n^{\H}\bm h_n\right\|_{\F}^2}, \forall n
\end{equation}
with given $t$.

Substituting \eqref{eq:sb:01} and \eqref{eq:sb:06} into problem \eqref{eq:26}, we obtain a downgraded problem as
\begin{subequations}\label{eq:sb:08}
\begin{align}
\min\limits_{\scriptstyle{\overline{\bm Q}, \{\eta_i\}_{i \in \cal I}}\hfill\atop
\scriptstyle{\left\{ {{\zeta _{m,n}}} \right\}_{m \in {\cal M},n \in {\cal N}}}} &\;\Tr\br{\overline{\bm Q}}\\
\st&\; \zeta_{m, n} \ge 0, \overline{\bm Q}  \succeq \bm 0, \forall m, n\\
&\; \eta_i \ge 0, \forall i\\
&\; \bm\Phi_{m,n}\br{\widehat{\bm X}_n, \zeta_{m,n}} \succeq \bm 0, \forall m, n \\
&\; \bm \Psi_i\br{\widehat{\bm Y}, \eta_i} \succeq \bm 0, \forall i
\end{align}
\end{subequations}
where
\begin{multline}\label{eq:sb:09}
\widehat{\bm X}_n = \frac{p_n}{1 - \exp\br{-\widetilde{R}^{\REQ}_{e, m\rightarrow n}}}\frac{\bm\Xi_n\bm\Xi_n^{\H}\bm H_n\bm\Xi_n\bm\Xi_n^{\H}}{\left\|\bm\Xi_n^{\H}\bm h_n\right\|_{\F}^2}
\\
- \sum\limits_{k=1}^N p_k\frac{\bm\Xi_k\bm\Xi_k^{\H}\bm H_k\bm\Xi_k\bm\Xi_k^{\H}}{\left\|\bm\Xi_k^{\H}\bm h_k\right\|_{\F}^2}
- \bm\Phi\overline{\bm Q}\bm\Phi^{\H}, \forall n
\end{multline}
and
\begin{equation}\label{eq:sb:10}
\widehat{\bm Y} = \sum\limits_{n=1}^N p_n\frac{\bm\Xi_n\bm\Xi_n^{\H}\bm H_n\bm\Xi_n\bm\Xi_n^{\H}}{\left\|\bm\Xi_n^{\H}\bm h_n\right\|_{\F}^2}
- \bm\Phi\overline{\bm Q}\bm\Phi^{\H}.
\end{equation}

Solving the optimization problem \eqref{eq:sb:08} in the third step of the SDP-TsBAJ algorithm, we obtain the MRT-ZFBF-TsBAJ algorithm.

{\color{black}
\emph{Complexity Comparison:}
The optimization problem \eqref{eq:26} has $N+1$ matrix variables and $MN+I$ real variables, and each matrix variable is of size $N_t \times N_t$.
Therefore, the number of decision variables of the SDP-TsBAJ algorithm is at the order of $n_2 = \br{N+1}N_t^2 + MN + I$.
Moreover, the feasible region of problem \eqref{eq:26} contains $N+1$ positive-semidefinite constraints with size $N_t \times N_t$, $MN+I$ positive-semidefinite constraints with size $\br{N_t + 1} \times \br{N_t + 1}$ and $MN+I+N$ linear constraints.
Hence, the number of iterations is at order of $n_1 = \log\epsilon^{-1}\sqrt{\br{N+1}N_t + \br{MN+I}\br{N_t + 2} +N}$.
The overall computational complexity is calculated as $\OO{n_1\br{n_2m_1 + n_2^2 m_2 + n_2^3}}$, where $m_1 = \br{N+1}N_t^3 + \br{MN+I}\br{N_t+1}^3 + MN+I+N$ and $m_2 = \br{N+1}N_t^2 + \br{MN+I}\br{N_t+1}^2 + MN+I+N$.
Here, the operator $\cal O$ is the worst case computation bound \cite{Ben-Tal2001}.
Following the similar procedure, we summarize the computational complexities of the proposed SDP-TsBAJ, ZFBF-TsBAJ and MRT-ZFBF-TsBAJ in Table \ref{tb:cplex} with $T$ as the number of one-dimensional search.

\begin{table*}[htbp]\tiny
  \caption{Computational Complexity, $\OO{T n_1\br{n_2m_1 + n_2^2m_2 + n_2^3}}$}\label{tb:cplex}
  \centering
  \begin{tabular}{|l|l|}
    \hline
    \multicolumn{1}{|l|}{Algorithms} & Complexity Calculation \\\hline
    \multirow{4}[0]{*}{SDP-TsBAJ} & $n_1 = \log\frac{1}{\epsilon}\sqrt{\br{N+1}N_t + \br{MN+I}\br{N_t + 2} +N}$ \\\cline{2-2}
    & $n_2 = \br{N+1}N_t^2 + MN + I$ \\\cline{2-2}
    & $m_1 = \br{N+1}N_t^3 + \br{MN+I}\br{N_t + 1}^3 + MN+I+N$ \\\cline{2-2}
    & $m_2 = \br{N+1}N_t^2 + \br{MN+I}\br{N_t + 2}^2 + MN+I+N$ \\\hline
    \multirow{4}[0]{*}{ZFBF-TsBAJ} &
    $n_1 = \log\frac{1}{\epsilon}\sqrt{N\br{N_t-N+1} + N_t + \br{MN+I}\br{N_t + 2}}$ \\\cline{2-2}
    & $n_2 = N\br{N_t-N+1}^2 + \br{N_t-N}^2 + MN+I$ \\\cline{2-2}
    & $m_1 = N\br{N_t-N+1}^3 + \br{N_t-N}^3 + \br{MN+I}\br{N_t + 1}^3 + MN+I+N$ \\\cline{2-2}
    & $m_2 = N\br{N_t-N+1}^2 + \br{N_t-N}^2 + \br{MN+I}\br{N_t + 1}^2 + MN+I+N$ \\\hline
    \multirow{4}[0]{*}{MRT-ZFBF-TsBAJ} &
    $n_1 = \log\frac{1}{\epsilon}\sqrt{N_t + \br{MN+I}\br{N_t + 2}}$ \\\cline{2-2}
    & $n_2 = \br{N_t-N}^2 + MN+I$ \\\cline{2-2}
    & $m_1 = \br{N_t-N}^3 + \br{MN+I}\br{N_t + 1}^3 + MN+I$ \\\cline{2-2}
    & $m_2 = \br{N_t-N}^2 + \br{MN+I}\br{N_t + 1}^2 + MN+I$ \\\hline
  \end{tabular}%
\end{table*}%

We observe that the computational complexity of the proposed SDP-TsBAJ, ZFBF-TsBAJ and MRT-ZFBF-TsBAJ algorithms scale linearly with the number of one-dimensional search points.
The SDP-TsBAJ algorithm has the highest computational complexity among the three proposed algorithms.
Then, the computational complexity is followed by the proposed ZFBF-TsBAJ algorithm.
The computational complexity of the MRT-ZFBF-TsBAJ algorithm is the lowest.
}

{\color{black}
\section{Numerical Results}
In this section, we present simulation results to demonstrate the performances of the proposed algorithms.~The parameters $\Omega_{u,n}$, $\Omega_{e,m}$ and $\Omega_{h,i}$ follow the setup of indoor channels as $17.3 + 24.9\log_{10}f_c + 38.3\log_{10} d$ dB, where $f_c$ and $d$ are respectively the carrier frequency and link distance \cite{3gpp_pathloss}.
The values of $\Theta^2_{e, m}$ and $\Theta^2_{h, i}$ are set as $0.05\Omega_{e,m}$ and $0.05\Omega_{h,i}$, respectively.
Unless otherwise specified, the simulation parameters are set in Table \ref{table:03}.

\begin{table*}[htbp]\tiny
  \caption{Simulation Parameters Setting}\label{table:03}
  \centering
  \begin{tabular}{|l|l|c|c|}
    \hline
    \multicolumn{2}{|l|}{\textbf{Parameters}} & \textbf{Symbols} & \textbf{Values} \\\hline
    \multicolumn{2}{|l|}{Carrier frequency and system bandwidth} & $f_c$, ${\cal BW}$ & 900 MHz, 200 KHz  \\\hline
    \multicolumn{2}{|l|}{Number of LUEs, EVEs and EHNs} & $N$, $M$, $I$ & 3, 2, 2 \\\hline
    \multicolumn{2}{|l|}{Number of antennas at BST} & $N_t$ & 7 \\\hline
    \multicolumn{2}{|l|}{Power of AWGN} & $\sigma_{n, u}^2$, $\sigma_{e, m}^2$, $\sigma_{h, i}^2$ & -30 dBm \\\hline
    \multicolumn{2}{|l|}{Maximum Tx power of BS} & $P^{\max}$ & 43 dBm \\\hline
    \multicolumn{2}{|l|}{Required harvested power of the $i$-th EHN} & $P_{h, i}^{\REQ}$ & -5 dBm \\\hline
    \multicolumn{2}{|l|}{Energy conversion efficiency of the $i$-th EHN} & $\xi_{h, i}$ & 0.8 \\\hline
    \multicolumn{2}{|l|}{Rate of auxiliary information of the $n$-th LUE} &  $R^{\REQ}_{e, n}$ & 100 Knats/sec \\\hline
    \multicolumn{2}{|l|}{Secrecy rate ratios} & $\varphi_1, \varphi_2, \varphi_3$ & 0.4, 0.3, 0.3\\\hline
    \multicolumn{2}{|l|}{Amplifier efficiency} & $\phi$ & 0.8 \\\hline
    \multicolumn{2}{|l|}{\color{black}{Distances of EVEs and EHNs to BST}} & \color{black}{$d_{e, m}$ and $d_{h, i}$} & \color{black}{8 m and 6 m} \\\hline
    \multicolumn{2}{|l|}{\color{black}{Distances of LUEs, EVEs and EHNs to BST}} & \color{black}{$d_{u, n}$} & \color{black}{16 m, 19 m and 22 m} \\\hline
  \end{tabular}%
\end{table*}%

\begin{figure}[htb]
\vspace{-0.2 cm}
\centering
  \includegraphics[width= 3 in]{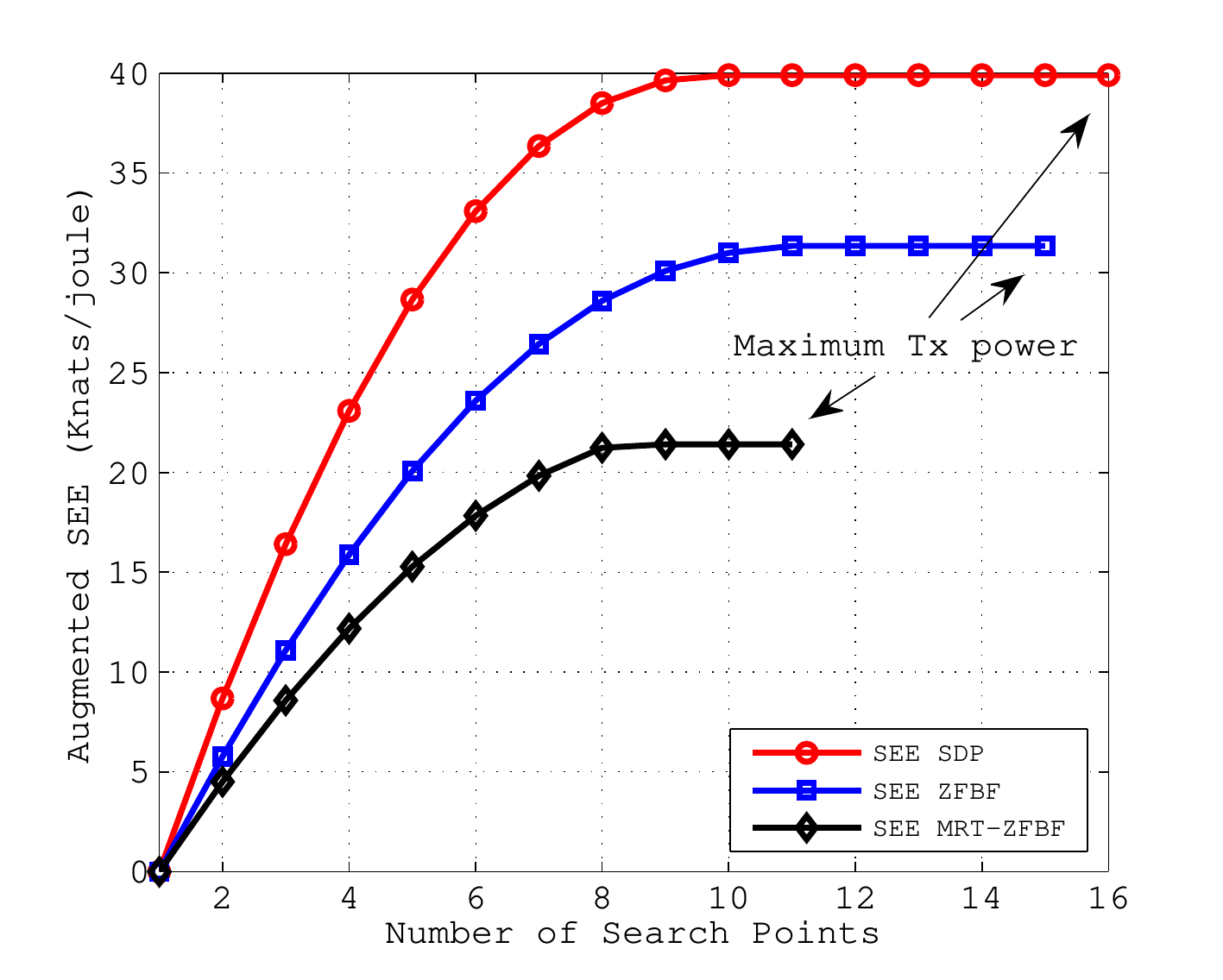}
  \vspace{-0.2 cm}
  \caption{The convergence of augmented SEE sequence.}\label{fg:sm:01a}
  \includegraphics[width= 3 in]{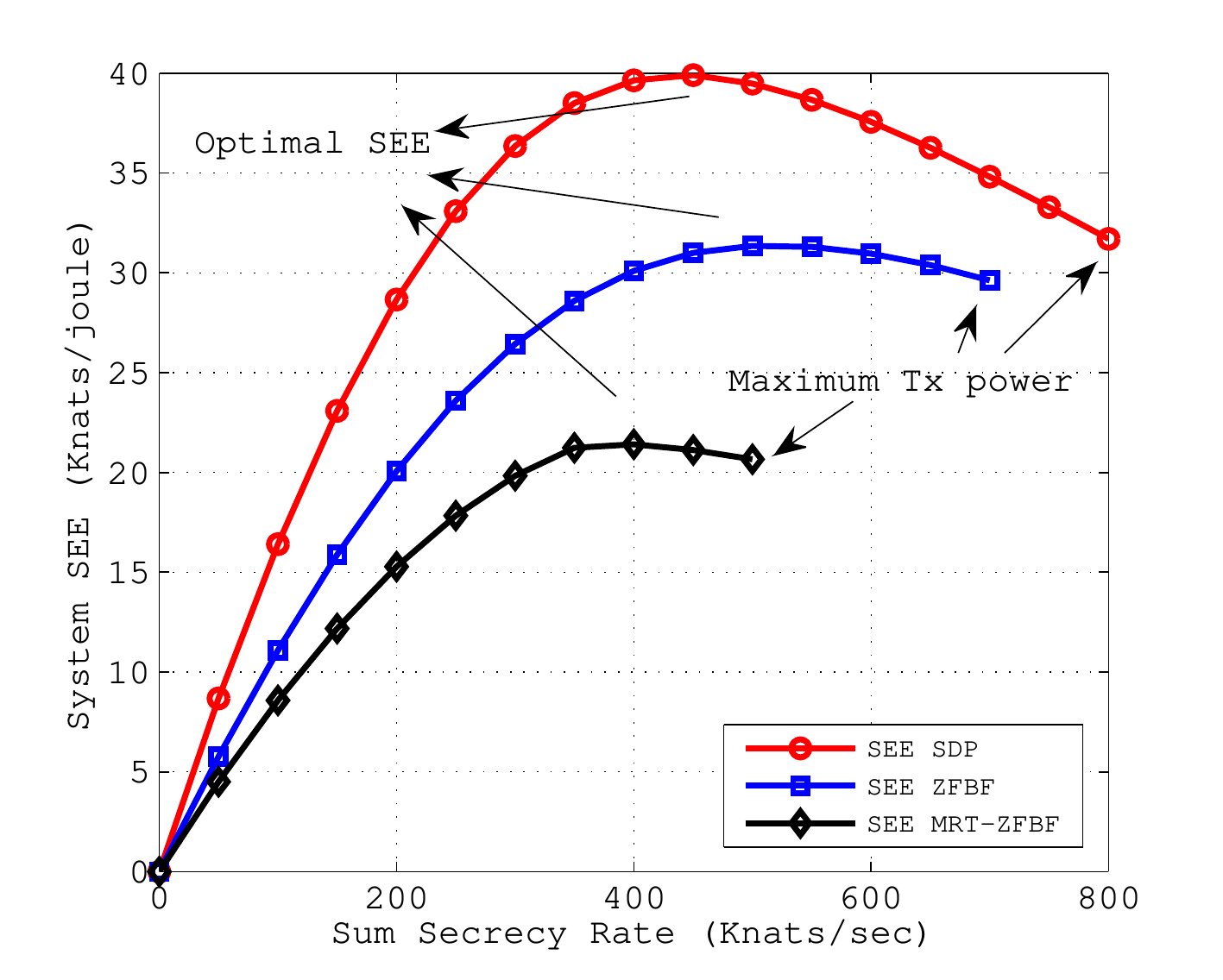}
  \vspace{-0.2 cm}
  \caption{System SEE varies with sum secrecy rate.}\label{fg:sm:01b}
  \vspace{-0.2 cm}
\end{figure}

Figure \ref{fg:sm:01a} shows the convergence behavior of the augmented SEE sequence for the proposed SDP-TsBAJ, ZFBF-TsBAJ and MRT-ZFBF-TsBAJ algorithms in a single channel implementation.
Since the augmented SEE sequence contains the optimal SEE with unit Knats/joule given the sum secrecy rate $t$, we observe that the proposed algorithms converge after at most $10$ searching attempts.
From Fig. \ref{fg:sm:01b}, we also observe that the system SEE increases first and starts to decrease after reaching the peak system SEE.
Therefore, we conclude that the proposed one-dimensional search is effective in solving the formulated SEE maximization problem.

\begin{figure}[htb]
\vspace{-0.2 cm}
\centering
  \includegraphics[width= 3 in]{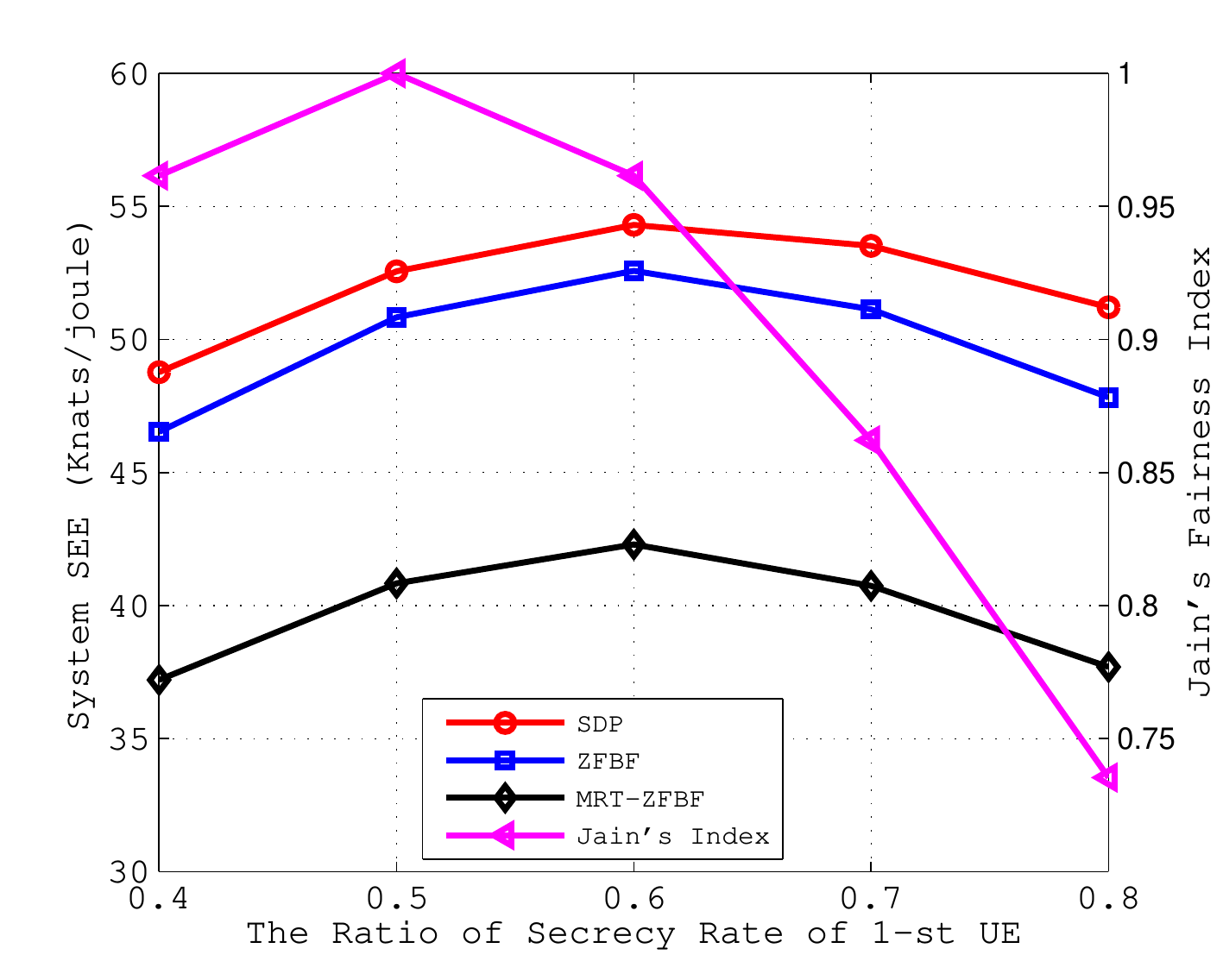}
  \vspace{-0.2 cm}
  \caption{The variation of Jain's fairness index and system SEE with ratio of secrecy rate of the $1$-st LUE in the two-LUE case.}\label{fg:sm:00}
  \vspace{-0.2 cm}
\end{figure}

Figure \ref{fg:sm:00} shows the variation of the Jain's fairness index and the system SEE with the secrecy-rate ratio of the first LUE in the two-LUE case.
Here, the Jain's fairness index is obtained as
\begin{equation}
\mbox{Jain's Index} =  \left({2 \sum\limits_{n=1}^2 \varphi_n^2}\right)^{-1}
\end{equation}
with $\sum\nolimits_{n=1}^2 \varphi_n = 1$. The value of Jain's fairness index takes value within the region $\sq{0.5, 1}$ in the two-LUE case.
A larger value of Jain's fairness index indicates that the LUEs are more satisfied  in the MISOME-SWIPT system.

In the two-LUE case, the distances of the first and second LUE to the BST are, respectively, set as $16$ m and $19$ m.
Hence, the first LUE is referred to as strong LUE, and the second LUE is weak LUE due to the larger pathloss compared with the first LUE. We observe that increasing the secrecy-rate ratio $\varphi_1$ induces the enhancement of both system SEE and Jain's fairness index when $\varphi_1 \le 0.5$.
This is due to the fact that enhancing the secrecy rate of the strong LUE reduces the secrecy-rate gap between the weak LUE and the strong LUE.
Moreover, it is easier to increase the secrecy rate of the strong LUE due to its smaller pathloss compared with that of the weak LUE.
When the value of $\varphi_1$ exceeds $0.5$, the secrecy-rate gap between the strong LUE and the weak LUE increases.
Therefore, the value of Jain's fairness index reduces, and the satisfactory level of LUEs decreases.
However, we observe that the system SEE increases with the value of $\varphi_1$ when $\varphi_1 \in \sq{0.5, 0.6}$.
After reaching the peak value with $\varphi_1 = 0.6$, the system SEE decreases.
This observation is due to the fact that an increased value of  secrecy-rate ratio reduces the flexibility in suppressing the EVEs and powering the EHNs.
Therefore, we conclude that the system SEE and fairness can be balanced with a proper value of $\varphi_1 \in \sq{0.5, 0.6}$.

\begin{figure}[htb]
\vspace{-0.2 cm}
\centering
  \includegraphics[width= 3 in]{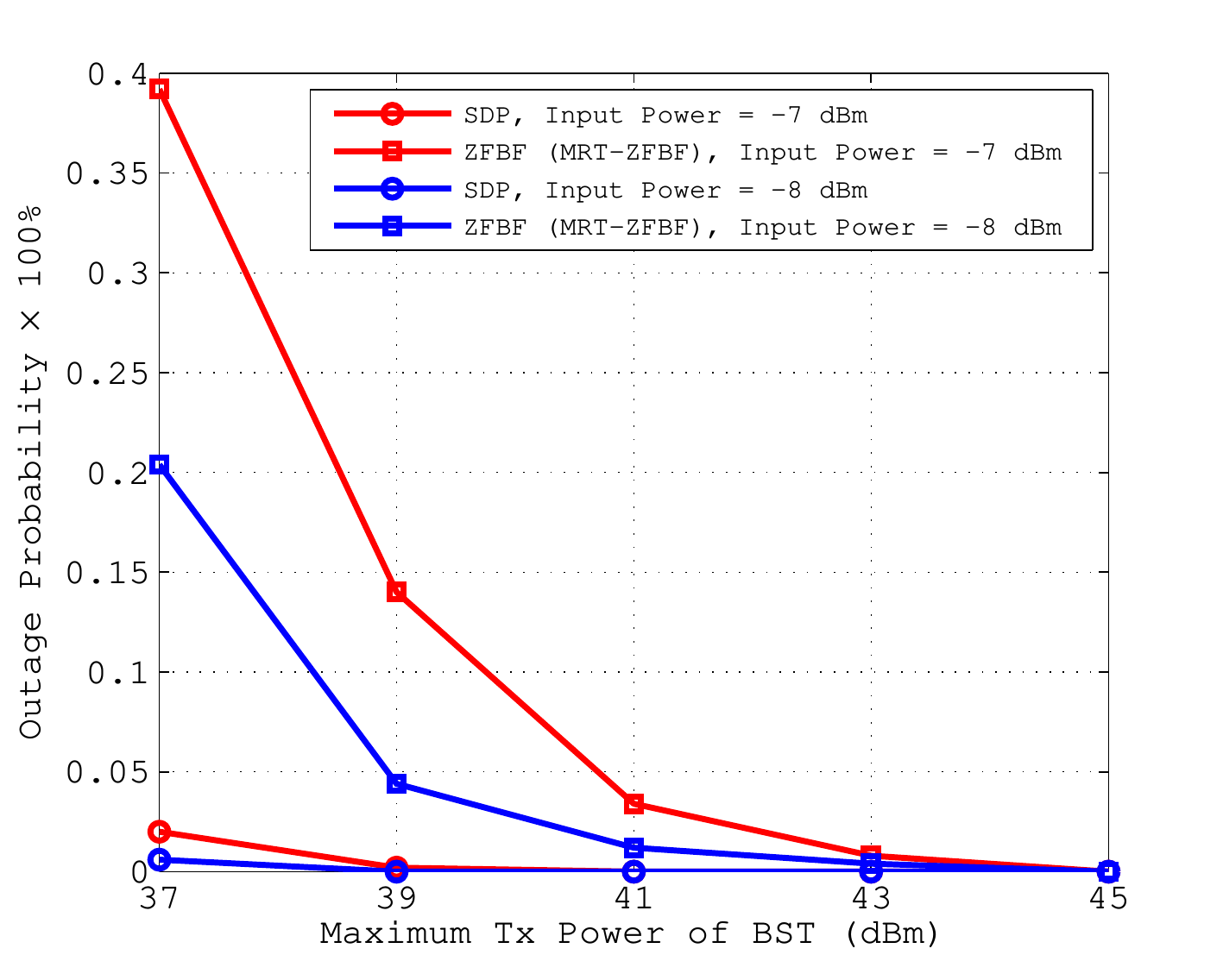}
  \vspace{-0.2 cm}
  \caption{The variation of outage probability with maximum transmit power of BST.}\label{fg:sm:02}
  \vspace{-0.2 cm}
\end{figure}

Figure \ref{fg:sm:02} shows that the outage probability of MISOME-SWIPT system varies with the maximum transmit power of BST.
Here, the outage event of the MISOME-SWIPT system is defined as the minimum harvested energy is below the activation threshold of energy harvesters of EHNs.
The proposed ZFBF-TsBAJ and MRT-ZFBF-TsBAJ algorithms share the same outage probability since both  ZFBF-TsBAJ and MRT-ZFBF-TsBAJ algorithms share the same feasible region.
We observe that the outage probability of MISOME-SWIPT system monotonically decreases with the maximum transmit power of BST.
Moreover, the outage probability approaches zero and the energy harvesters of EHNs are activated with high probability when the maximum transmit power of BST exceeds $43$ dBm.

In order to demonstrate the effectiveness of the proposed algorithms, we consider another set of baseline schemes in the remaining simulations, where the BST solves the SRM problem with PSR constraints.
Hereinafter, we name the considered  baselines as SRM-SDP, SRM-ZFBF and SRM-MRT-ZFBF algorithms.
In addition, when $N_t = 7$, $N = 3$, $M = 2$ and $I = 2$, the number of arithmetical manipulations for SDP-TsBAJ, ZFBF-TsBAJ and MRT-ZFBF-TsBAJ algorithms are, respectively, $1.1678\times 10^9$, $7.9845\times 10^8$ and $6.2297\times 10^7$. We observe that the computational complexities of ZFBF-TsBAJ and MRT-ZFBF-TsBAJ algorithms are only 68.37\% and 5.33\% of that of SDP-TsBAJ algorithms.

\begin{figure}[htb]
\vspace{-0.2 cm}
\centering
  \includegraphics[width= 3 in]{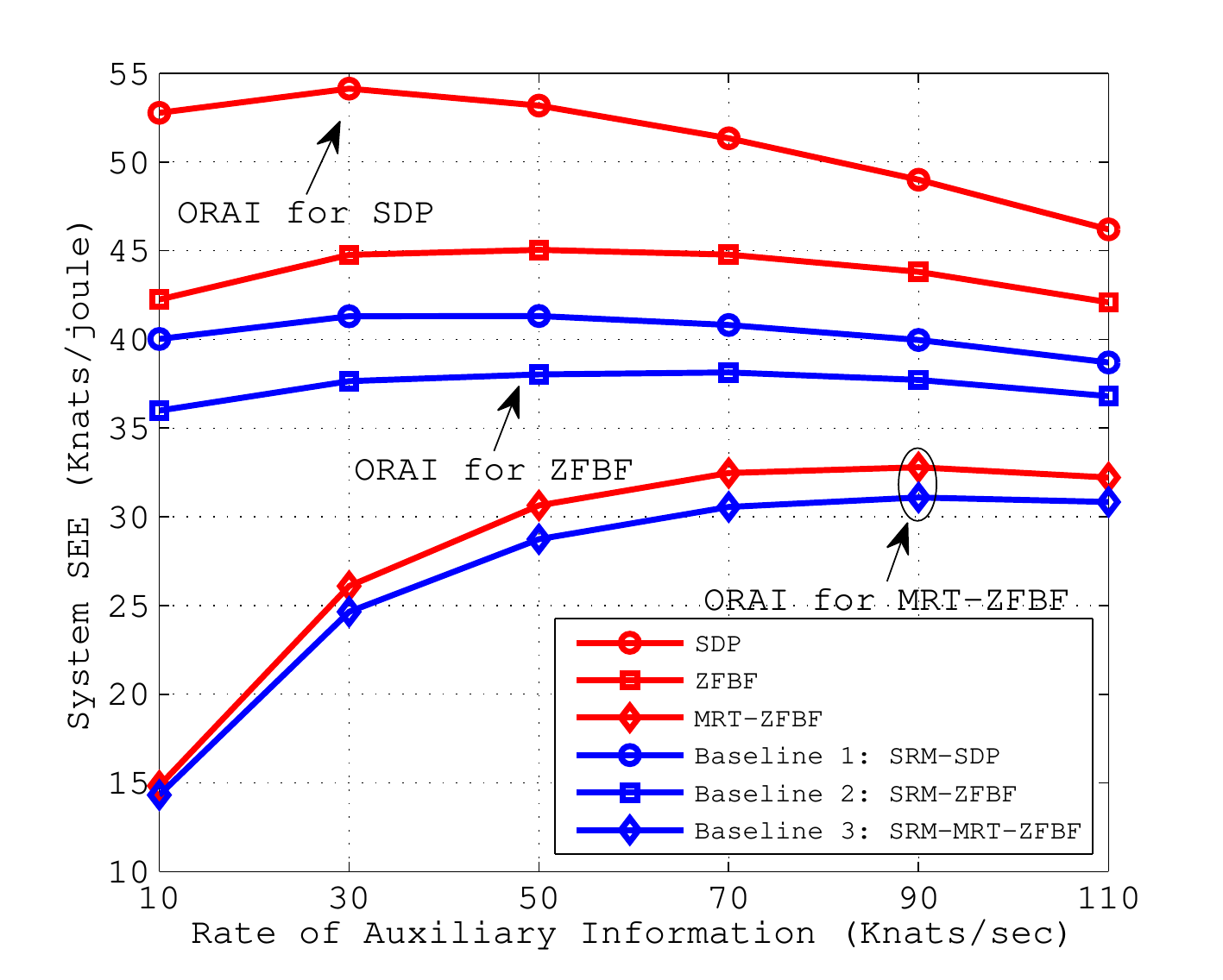}
    \vspace{-0.2 cm}
  \caption{System SEE v.s. rate of auxiliary information.}\label{fg:sm:06a}
  \includegraphics[width= 3 in]{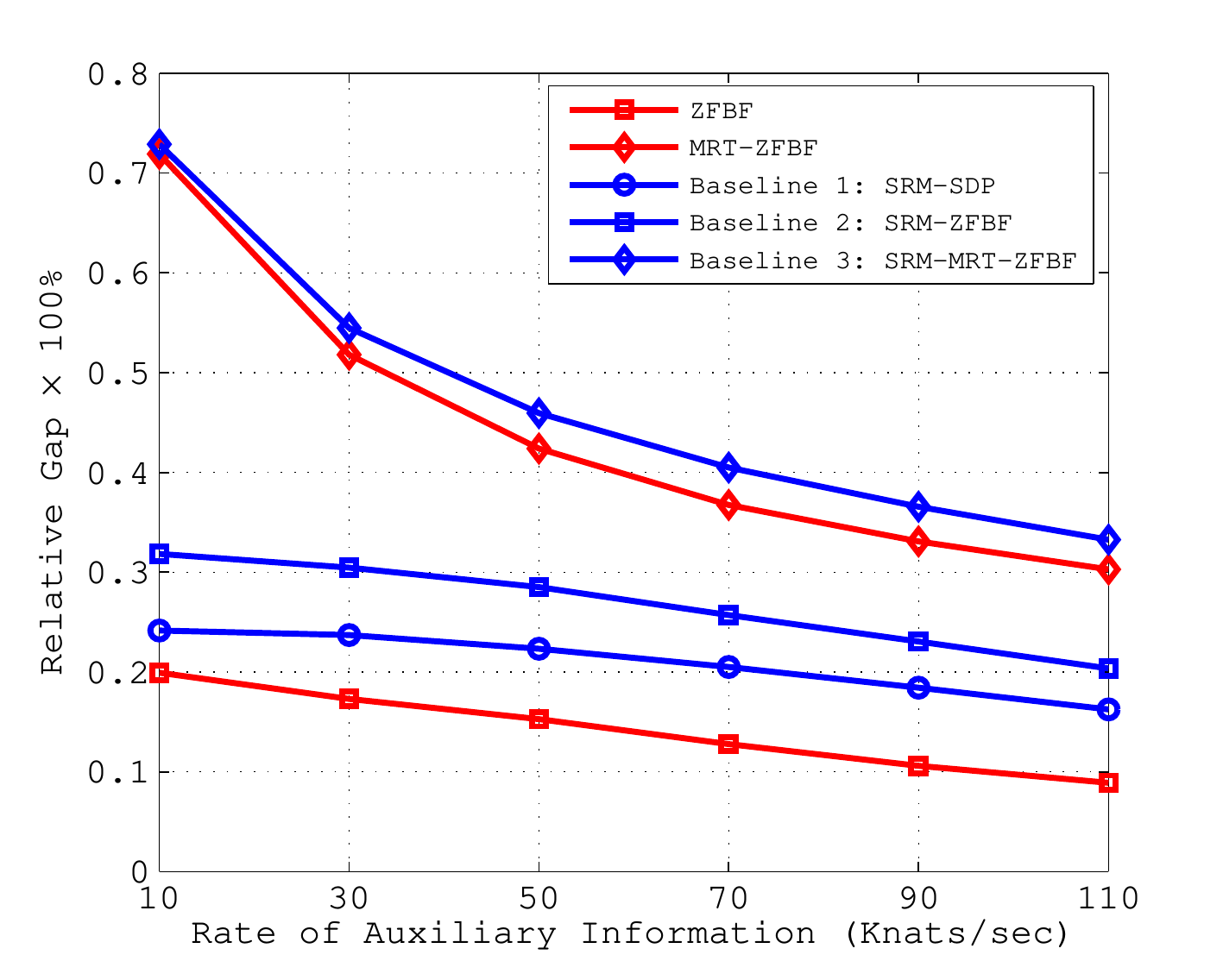}
    \vspace{-0.2 cm}
  \caption{Relative gap v.s. rate of auxiliary information.}\label{fg:sm:06b}
  \vspace{-0.2 cm}
\end{figure}

Figure \ref{fg:sm:06a} shows that the system SEE varies with the rate of auxiliary information, and Fig. \ref{fg:sm:06b} shows the relative gaps between the SDP-TsBAJ algorithm with two suboptimal algorithms (ZFBF-TsBAJ and MRT-ZFBF-TsBAJ algorithms) and three baseline algorithms (SRM-SDP, SRM-ZFBF and SRM-MRT-ZFBF algorithms).
Here, the relative gap is obtained as
\begin{equation}
\mbox{Relative Gap} = \frac{\mbox{SEE}_{\rm SDP-TsBAJ} - \mbox{SEE}_{\pi}}{\mbox{SEE}_{\rm SDP-TsBAJ}}
\end{equation}
where $\pi$ can be ZFBF-TsBAJ, MRT-ZFBF-TsBAJ, SRM-SDP, SRM-ZFBF and SRM-MRT-ZFBF algorithms.

We observe that the system SEE increases monotonically with the rate of auxiliary information.
After reaching the peak, the system SEE starts to decreases with the rate of auxiliary information.
This observation can be explained as follows: 1) a smaller rate of auxiliary information ($\le 30$ Knats/sec for SDP-TsBAJ, $\le 50$ Knats/sec for ZFBF-TsBAJ and $\le 90$ Knats/sec for MRT-ZFBF-TsBAJ) restricts the flexibility in enhancing the sum secrecy rate; and 2) larger rate of auxiliary information ($> 30$ Knats/sec for SDP-TsBAJ, $> 50$ Knats/sec for ZFBF-TsBAJ and $> 90$ Knats/sec for MRT-ZFBF-TsBAJ) sacrifices the rate of main information.
Therefore, we conclude that there exists an optimal rate of auxiliary information (ORAI).
For example, the ORAI for SDP based algorithms (SDP-TsBAJ and SRM-SDP algorithms) is around $30$ Knats/sec as shown in Fig. \ref{fg:sm:06a}.
The ORAIs for the ZFBF based algorithms (ZFBF-TsBAJ and SRM-ZFBF algorithms) and MRT-ZFBF based algorithms (MRT-ZFBF-TsBAJ and SRM-MRT-ZFBF algorithms) are $50$ Knats/sec and $90$ Knats/sec as shown in Fig. \ref{fg:sm:06a}.

From Fig. \ref{fg:sm:06b}, we observe that the SEE performance of ZFBF-TsBAJ and MRT-ZFBF-TsBAJ algorithms approaches that of the SDP-TsBAJ algorithm when the rate of auxiliary information increases.
Moreover, the relative gap between the SDP-TsBAJ and MRT-ZFBF-TsBAJ decreases rapidly  with the rate of auxiliary information (less than 30\% when $R^{\REQ}_{e, n} = 110$ Knats/sec).
Hence, the MRT-ZFBF-TsBAJ algorithm is also a feasible solution since the computational complexity of MRT-ZFBF-TsBAJ algorithm is 5\% of the SDP-TsBAJ algorithm.

\begin{figure}[htb]
\vspace{-0.2 cm}
\centering
  \includegraphics[width= 3 in]{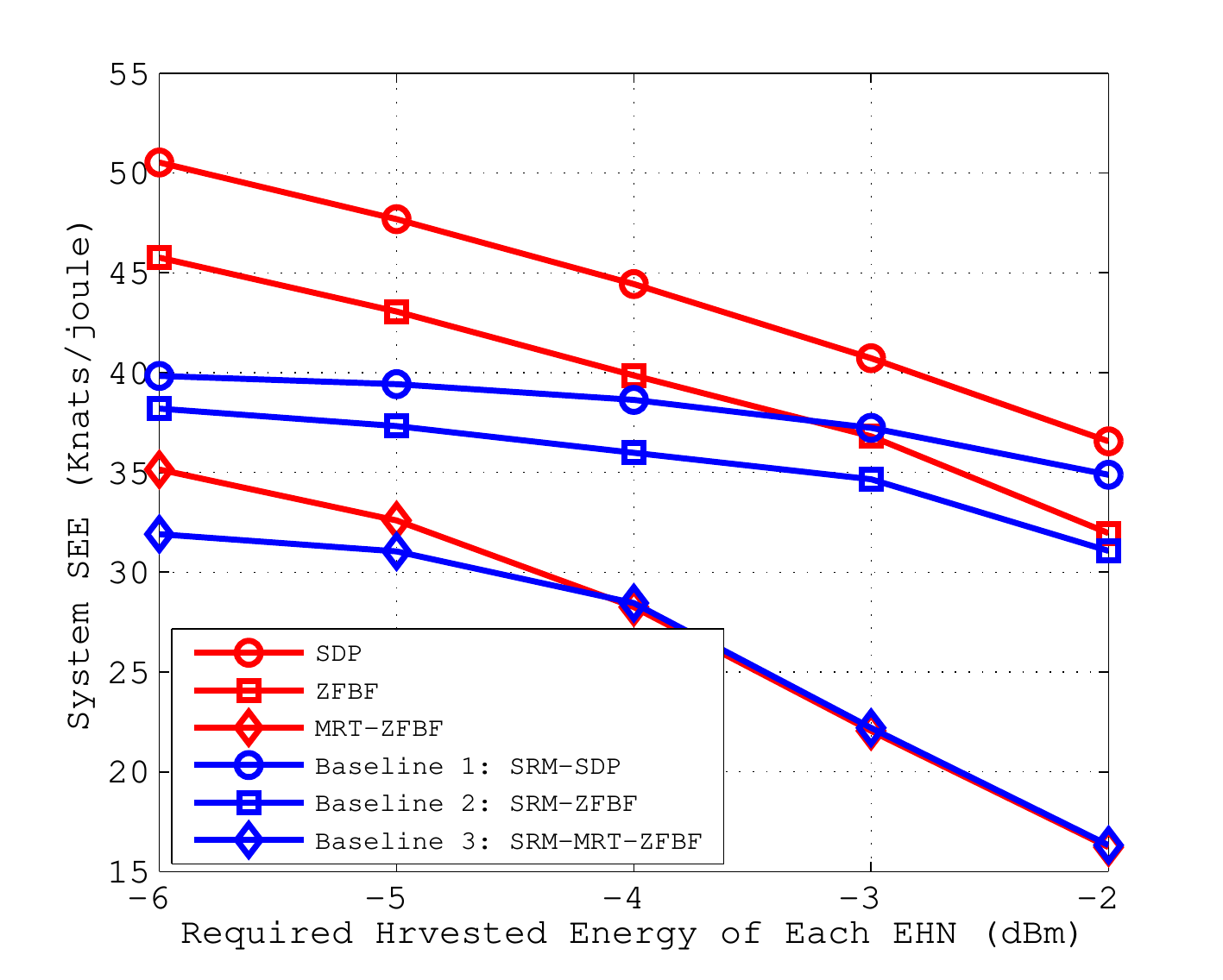}
  \vspace{-0.2 cm}
  \caption{The variation of the system SEE with requirement of harvested energy.}\label{fg:sm:03a}
  \vspace{-0.2 cm}
\end{figure}

Figure \ref{fg:sm:03a} shows the variation of system SEE with minimum harvested energy.
We observe that the system SEE decreases with minimum harvested energy for the SDP-TsBAJ, ZFBF-TsBAJ algorithm and MRT-ZFBF-TsBAJ algorithms as well as three baseline algorithms.
This is a result of two facts: 1) a higher RF-EH demand leads to a higher power consumption at BST; and 2) the system secrecy rate becomes smaller with the higher RF-EH demand.
From Fig. \ref{fg:sm:03a}, we also observe that performance gap between SDP-TsBAJ and SRM-SDP diminishes with the increase in minimum harvested energy.
This is due to the fact that a higher RF-EH demand consumes more power on the energy delivery to EHNs such that the probability of BST draining its power budget becomes higher.
Therefore, the SEE maximization problem is reduced  to the SRM maximization problem when the minimum harvested energy increases.

\section{Conclusion}
We developed the TsBAJ framework for the SEE maximization via the joint design of beamforming vector and AN covariance matrix in the MISOME-SWIPT systems, where the CSI of EVEs and EHNs is imperfect.
We considered the PSR constraints to guarantee the fairness among LUEs.
Due to the non-convexity in the PSR constraints and SEE function as well as the infinite amount of RF-EH constraints and information-leakage constraints, we exploited the structure of formulated problem to propose efficient algorithms.
Leveraging the SDR technique and one-dimensional search, we obtained the optimal solution within the TsBAJ framework.
Moreover, we also developed two suboptimal solutions based on the heuristic beamforming techniques: ZFBF and MRT.
Simulation results show that the system SEE of proposed SDP-TsBAJ algorithm outperforms the benchmarks, and the introduced PSR constraints can be used to strike the balance between the fairness and system SEE.
Moreover, simulation results also demonstrate that the ZFBF-TsBAJ algorithm can approach the performance of SDP-TsBAJ algorithm when the rate of auxiliary information is large.
}

\appendices
{\color{black}
\section{Proof of Proposition \ref{pr:01}}\label{apdx:01}
Suppose $\br{\left\{ {\bm W_n^*} \right\}_{n \in {\cal N}}, \bm Q^*, t^*}$ is a solution to the problem \eqref{eq:21} such that the inequalities in \eqref{eq:19} and \eqref{eq:20} are inactive.
Therefore, we can find a set of scale factors $\{y_n\}_{n \in \cal N}$ with $y_n \in \sq{0, 1}$.

Scaling the beamforming matrices $\{\bm W_n^*\}_{n \in \cal N}$ via the scale factors $\{y_n\}_{n \in \cal N}$, we can construct another set of beamforming matrices $\{\widehat{\bm W}_n^*\}_{n \in \cal N}$ and AN covariance matrix $\widehat{\bm Q}^*$ as
\begin{subequations}
\begin{align}
\widehat{\bm W}_n^* &= y_n \bm W_n^* \succeq \bm 0, \forall n \label{eq:apdx:01}\\
\widehat{\bm Q}^* &= {\bm Q}^* + \sumn \br{1-y_n}\bm W_n^* \succeq \bm 0 \label{eq:apdx:02}
\end{align}
\end{subequations}
such that the inequalities in \eqref{eq:19} and \eqref{eq:20} are active.

Substituting \eqref{eq:apdx:01} and \eqref{eq:apdx:02} into the inequalities in \eqref{eq:20}, we obtain the value of $y_n$ as
\begin{equation}\label{eq:apdx:03}
y_n = \frac{\theta_n\br{t^*}}{1 + \theta_n\br{t^*}}\frac{\Tr\br{\bm H_n\br{\sumn \bm W_n^* + \bm Q^*}} + \sigma_{u,n}^2}{\Tr\br{\bm H_n \bm W_n^*}}
\end{equation}
such that the inequalities in \eqref{eq:20} become active.~Since the equalities in \eqref{eq:20} hold, the inequality in \eqref{eq:19} is active.~Moreover, we observe that the constructed beamforming matrices $\{\widehat{\bm W}_n^*\}_{n \in \cal N}$ and AN covariance matrix $\widehat{\bm Q}^*$ also satisfy the constraints in \eqref{eq:15c}-\eqref{eq:15g} and \eqref{eq:20}.~Hence, we conclude that the constructed optimal solution $(\{ \widehat{\bm W}_n^* \}_{n \in {\cal N}}, \widehat{\bm Q}^*, t^*)$ to problem \eqref{eq:21} guarantees that the inequalities in \eqref{eq:19} and \eqref{eq:20} are active.

\section{Proof of Equivalence Between \eqref{eq:15} and \eqref{eq:21}}\label{apdx:01a}
Assuming that $\br{\{\bm W_n^*\}_{n \in {\cal N}}, \bm Q^*}$ is an optimal solution to \eqref{eq:15}, we can construct an optimal solution to problem \eqref{eq:21} by letting $t^* = \sum\nolimits_{n=1}^N R_{u,n}^{\SEC}\br{\{\bm W_n^*\}_{n \in {\cal N}}, \bm Q^*}$.
Substituting $\br{\{\bm W_n^*\}_{n \in {\cal N}}, \bm Q^*, t^*}$ into problem \eqref{eq:21}, we can verify that the constraints in \eqref{eq:21b} are satisfied, and the objective value of problem \eqref{eq:15} is also equal to that of problem \eqref{eq:21}.
Hence, we conclude that $\br{\{\bm W_n^*\}_{n \in {\cal N}}, \bm Q^*, t^*}$ is an optimal solution to problem \eqref{eq:21}.
When $\br{\{\bm W_n^*\}_{n \in {\cal N}}, \bm Q^*, t^*}$ is an optimal solution to \eqref{eq:21}, we consider two cases as follows:
\begin{itemize}
  \item if $\br{\{\bm W_n^*\}_{n \in {\cal N}}, \bm Q^*, t^*}$ can guarantee the activeness of \eqref{eq:19} and \eqref{eq:20}, $\br{\{\bm W_n^*\}_{n \in {\cal N}}, \bm Q^*}$ is an optimal solution to \eqref{eq:15}.
  \item if $\br{\{\bm W_n^*\}_{n \in {\cal N}}, \bm Q^*, t^*}$ \emph{cannot} guarantee the activeness of \eqref{eq:19} and \eqref{eq:20}, we can construct another optimal solution  to \eqref{eq:21}, $(\{ \widehat{\bm W}_n^* \}_{n \in {\cal N}}, \widehat{\bm Q}^*, t^*)$, as in \eqref{eq:pr1:01} such that the inequalities in \eqref{eq:19} and \eqref{eq:20} are active.
\end{itemize}

Substituting $(\{\widehat{\bm W}_n^* \}_{n \in {\cal N}}, \widehat{\bm Q}^*)$ into \eqref{eq:15}, the constraints in \eqref{eq:15b}--\eqref{eq:15g} are satisfied and the optimal value of \eqref{eq:15a} is equal to that of the optimal value of \eqref{eq:21a}.

Hence, we obtain the equivalence between the problems \eqref{eq:15} and \eqref{eq:21}.

}

\begin{figure*}[ht]
\begin{align}
{\OPT}\br{t_1} =& L\br{{t_2}, \{\bm W_n^{t_2}, \alpha_n^{t_2}\}_{n \in \cal N}, \bm Q^{t_2}, \{\eta_i^{t_2},  \bm \Gamma_i^{t_2}\}_{i \in \cal I}, \{\zeta_{m, n}^{t_2}, \bm\Lambda_{m, n}^{t_2}\}_{{m\in\cal M}, {n\in\cal N}}} \label{eq:apdx02:02}\\
{\mathop\ge\limits^{(a)}}& L\br{{t_2}, \{\bm W_n^{t_2}, \alpha_n^{t_1}\}_{n \in \cal N}, \bm Q^{t_2}, \{\eta_i^{t_2}, \bm \Gamma_i^{t_1}\}_{i \in \cal I}, \{\zeta_{m, n}^{t_2}, \bm\Lambda_{m, n}^{t_1}\}_{{m\in\cal M}, {n\in\cal N}}} \nonumber\\
{\mathop\ge\limits^{(b)}}& L\br{{t_1}, \{\bm W_n^{t_2}, \alpha_n^{t_1}\}_{n \in \cal N}, \bm Q^{t_2}, \{\eta_i^{t_2}, \bm \Gamma_i^{t_1}\}_{i \in \cal I}, \{\zeta_{m, n}^{t_2}, \bm\Lambda_{m, n}^{t_1}\}_{{m\in\cal M}, {n\in\cal N}}} \nonumber\\
{\mathop\ge\limits^{(c)}}& L\br{{t_1}, \{\bm W_n^{t_1}, \alpha_n^{t_1}\}_{n \in \cal N}, \bm Q^{t_1}, \{\eta_i^{t_1}, \bm \Gamma_i^{t_1}\}_{i \in \cal I}, \{\zeta_{m, n}^{t_1}, \bm\Lambda_{m, n}^{t_1}\}_{{m\in\cal M}, {n\in\cal N}}} = {\OPT}\br{t_1} \nonumber
\end{align}
\begin{align}
&\{\alpha_n^{t_2}\}_{n \in \cal N}, \{\bm \Gamma_i^{t_2}\}_{i \in \cal I} \mbox{ and } \{\bm\Lambda_{m, n}^{t_2}\}_{{m\in\cal M}, {n\in\cal N}} \label{eq:apdx02:03}\\
=& \mathop{\argmax}\limits_{\scriptstyle\{\alpha_n\}_{n \in\cal N}, \{\bm \Gamma_i\}_{i \in \cal I}, \hfill\atop
\scriptstyle\{\bm\Lambda_{m, n}\}_{{m\in\cal M}, {n\in\cal N}}} L\br{{t_2}, \{\bm W_n^{t_2}, \alpha_n\}_{n \in \cal N}, \bm Q^{t_2}, \{y_i^{t_2}, \bm \Gamma_i\}_{i \in \cal I}, \{\zeta_{m, n}^{t_2}, \bm\Lambda_{m, n}\}_{{m\in\cal M}, {n\in\cal N}}} \nonumber
\end{align}
\begin{align}
&\{\bm W_n^{t_1}\}_{n \in \cal N}, \bm Q^{t_1}, \{\eta_i^{t_1}\}_{i \in \cal I} \mbox{ and } \{\zeta_{m, n}\}_{{m\in\cal M}, {n\in\cal N}}  \label{eq:apdx02:04}\\
=& \mathop{\argmin}\limits_{\scriptstyle{\left\{ {\bm W_n} \right\}_{n \in {\cal N}}}, \bm Q, \{\eta_{i}\}_{i \in \cal I}\hfill\atop
\scriptstyle{\left\{ {{\zeta _{m,n}}} \right\}_{m \in {\cal M},n \in {\cal N}}}\hfill} L\br{{t_1}, \{\bm W_n, \alpha_n^{t_1}\}_{n \in \cal N}, \bm Q, \{\eta_i, \bm \Gamma_i^{t_1}\}_{i \in \cal I}, \{\zeta_{m, n}, \bm\Lambda_{m, n}^{t_1}\}_{{m\in\cal M}, {n\in\cal N}}}. \nonumber
\end{align}
\hrulefill
\vspace{-0.3 cm}
\end{figure*}

\section{Proof of Proposition \ref{pr:02}}\label{apdx:02}
Since the strong duality hold for problem \eqref{eq:26}, we have the optimal value of \eqref{eq:26} given $t$ as
\begin{align}
&{\OPT}\br{t} \label{eq:apdx02:01}\\
=& L\br{t, \{\bm W_n^t, \alpha_n^t\}_{n \in \cal N}, \bm Q^t, \{\eta_i^t, \bm \Gamma_i^t\}_{i \in \cal I}, \{\zeta_{m, n}^t, \bm\Lambda_{m, n}^t\}_{{m\in\cal M}, {n\in\cal N}}} \nonumber
\end{align}
where $\{\alpha_n^t\}_{n \in \cal N}$, $\{\bm \Gamma_i^t\}_{i \in \cal I}$ and $\{\bm\Lambda_{m, n}^t\}_{{m\in\cal M}, {n\in\cal N}}$ are the optimal dual variables given $t$.

Let $\{\alpha_n^{t_2}\}_{n \in \cal N}$, $\{\bm \Gamma_i^{t_2}\}_{i \in \cal I}$ and $\{\bm\Lambda_{m, n}^{t_2}\}_{{m\in\cal M}, {n\in\cal N}}$ denote the optimal dual variables, and $\{\bm W_n^{t_2}\}_{n \in \cal N}$, $\bm Q^{t_2}$, $\{\eta_i^{t_2}\}_{i \in \cal I}$ and $\{\zeta_{m, n}^{t_2}\}_{{m\in\cal M}, {n\in\cal N}}$ denote the optimal primal variables with respect to $t_2$.
Moreover, let $t_1 \le t_2$.
Then, we prove the monotonicity of ${\OPT}\br{t}$ based on following derivations

where step (a) is based on the fact in \eqref{eq:apdx02:03}; step (b) follows from fact that $\frac{1 + \theta_n\br{t}}{\theta_n\br{t}}$ is a monotonically decreasing function of $t$; and step (c) follows from \eqref{eq:apdx02:04}.

\section{Proof of Proposition \ref{pr:03}}\label{apdx:03}
Since the proof of Proposition \ref{pr:03} follows similar arguments in \cite[Proposition 1]{Dong2018a}, we will only sketch the procedures.
The Lagrangian of optimization problem \eqref{eq:26} is denoted as
\begin{align}
&L\br{\{\bm W_n, \alpha_n\}_{n \in \cal N}, \bm Q, \{\eta_i, \bm \Gamma_i\}_{i \in \cal I}, \{\zeta_{m, n}, \bm\Lambda_{m, n}\}_{{m\in\cal M}, {n\in\cal N}}} \nonumber\\
=& \sumn\Tr\br{\bm W_n\bm C_n } + \Tr\br{\bm Q\bm D} + \Upsilon \label{eq:apdx03:01}
\end{align}
where
$\bm C_n = \bm I_{N_t} + \sum\nolimits_{k=1}^N\alpha_k \bm H_k - \sum\nolimits_{i=1}^I\widetilde{\bm G}_{h,i}^{\H}\bm \Gamma_i\widetilde{\bm G}_{h,i} + \sum\nolimits_{m=1}^M \frac{1}{1 - \exp\br{-\widetilde{R}^{\REQ}_{e, m\rightarrow n}}}\widetilde{\bm G}_{e, m}^{\H}\bm\Lambda_{m, n}\widetilde{\bm G}_{e, m} + \sum\nolimits_{m=1}^M \sum\nolimits_{k=1}^N \widetilde{\bm G}_{e, m}^{\H}\bm\Lambda_{m, k}\widetilde{\bm G}_{e, m}- \alpha_n\frac{1 + \theta_n\br{t}}{\theta_n\br{t}}\bm H_n$,
$\bm D = \bm I_{N_t} + \sum\nolimits_{n=1}^N\alpha_n\bm H_n - \sum\nolimits_{i=1}^I\widetilde{\bm G}_{h,i}^{\H}\bm \Gamma_i\widetilde{\bm G}_{h,i} - \sum\nolimits_{m=1}^M \sum\nolimits_{n=1}^N\widetilde{\bm G}_{e, m}^{\H}\bm\Lambda_{m, n}\widetilde{\bm G}_{e, m}$ and
$\Upsilon = \sum\nolimits_{n=1}^N\alpha_n\sigma_{u,n}^2
- \sum\nolimits_{i=1}^I\Tr(\bm \Gamma_i\diag\{\eta_{i}\bm I_{N_t}, -\frac{P^{\REQ}_{h, i}}{\xi_{h, i}} - \eta_{i}\Theta_{h,i}^2\}) - \sum\nolimits_{m=1}^M \sum\nolimits_{n=1}^N\Tr\br{\bm\Lambda_{m, n}\diag\{\zeta_{m, n}\bm I_{N_t}, \sigma_{e,m}^2 - \zeta_{m, n}\Theta_{e,m}^2\}}$ with $n \in \cal N$.
Here, $\{\alpha_n\}_{n \in\cal N}$, $\{\bm \Gamma_i\}_{i \in \cal I}$ and $\{\bm\Lambda_{m, n}\}_{{m\in\cal M}, {n\in\cal N}}$ as the dual variables.

To guarantee the Lagrangian bounded below with $\br{\{\bm W_n^*\}, \bm Q^*, \{\eta_i^*\}_{i \in \cal I}, \{\zeta_{m, n}^*\}_{{m\in \cal M}, {n \in \cal N}}}$, the matrices $\bm C_n$ and $\bm D$ need to be positive-semidefinite, i.e.,
$\bm C_n^* \succeq \bm 0 \mbox{ and } \bm D^* \succeq \bm 0$, $n \in \cal N$.
Moreover, the Karush-Khun-Tucker (KKT) conditions related to $\bm W_n$ and $\bm Q$ are given as \begin{equation}\label{eq:apdx03:04a}
\bm W_n^*\bm C_n^* = \bm 0 \mbox{ and } \bm Q^*\bm D^* = \bm 0, \forall n.
\end{equation}

Define the matrix $\bm\Theta_n^*$ as
$\bm\Theta_n^* \triangleq \bm I_{N_t}
- \sum\nolimits_{i=1}^I\widetilde{\bm G}_{h,i}^{\H}\bm \Gamma_i^*\widetilde{\bm G}_{h,i}
+ \sum\nolimits_{m=1}^M\sum\nolimits_{k=1}^N\widetilde{\bm G}_{e, m}^{\H}\bm\Lambda_{m, k}^*\widetilde{\bm G}_{e, m}
+ \sum\nolimits_{k=1}^N\alpha_k^* \bm H_k
+ \sum\nolimits_{m=1}^M \frac{1}{1 - \exp\br{-\widetilde{R}^{\REQ}_{e, m\rightarrow n}}}\widetilde{\bm G}_{e, m}^{\H}\bm\Lambda_{m, n}^*\widetilde{\bm G}_{e, m}
 + \bm H_n$. Then, the optimal matrix $\bm C_n^*$ satisfies
\begin{equation}\label{eq:apdx03:08}
\bm C_n^* = \bm\Theta_n^* - \br{\alpha_n^*\frac{1 + \theta_n\br{t}}{\theta_n\br{t}} + 1}\bm H_n, \forall n.
\end{equation}

Now, we start to analyze the rank property of the matrix $\bm W_n^*$ as follows:
\begin{itemize}
  \item It is straightforward that the beamforming matrices satisfy the rank-one constraints when the matrix $\bm\Theta_n^*$ is full rank since $\Rank\br{\bm W_n^*} = \Rank\br{\bm \Theta_n^*\bm W_n^*} = \Rank\br{\bm H_n\bm W_n^*} \le 1$, $n \in \cal N$.
  \item If the matrix $\bm\Theta_n^*$ is rank-deficient, i.e., $\Rank\br{\bm\Theta_n^*} \le N_t$, we denote the rank of the matrix $\bm\Theta_n^*$ as $\Rank\br{\bm\Psi_n^*} = a_n$.
      Therefore, the rank of the null space of the matrix $\bm\Theta_n^*$ is $N_t - a_n$.
      Let the matrix $\bm \Pi_n = \sq{\bm\pi_{n, 1}, \bm\pi_{n, 2}, \ldots, \bm\pi_{n, N_t - a_n}}$ denote the orthogonal basis matrix of the null space of $\bm\Theta_n^*$.
      Since $\bm\pi_{n, k}^{\H}\bm\Theta_n^*\bm\pi_{n, k} = 0$, we have
      \begin{multline}\label{eq:apdx03:07}
      \bm\pi_{n, k}^{\H}\bm C_n^*\bm\pi_{n, k} = -\br{\alpha_n^*\frac{1 + \theta_n\br{t}}{\theta_n\br{t}} + 1}\\
      \times\bm\pi_{n, k}^{\H}\bm H_n \bm\pi_{n, k} \ge 0, \forall k, n
      \end{multline}
      where the inequality is based on the fact that $\bm C_n^*$ is positive-semidefinite.
      Here, the factor $\alpha_n^*\frac{1 + \theta_n\br{t}}{\theta_n\br{t}} + 1 > 0$.
      Since the condition in \eqref{eq:apdx03:07} holds with $\bm H_n = \bm h_n\bm h_n^{\H} \succeq \bm 0$, we have $\bm h_n^{\H} \bm\pi_{n, k} = 0$.
      Hence, the matrix $\bm\Pi_n$ is also the orthogonal basis matrix of the null space of $\bm H_n$ and $\bm C_n^*$.
      Therefore, $\Rank\br{\bm C_n^*} \le a_n$.
      Based on \eqref{eq:apdx03:08}, we also have $\Rank\br{\bm C_n^*} \ge a_n - 1$.
      We prove that $\Rank\br{\bm C_n^*} = a_n - 1$ via contradiction.
      If $\Rank\br{\bm C_n^*} = a_n$, the optimal matrix $\bm C_n^*$ shares the same null space with $\bm\Psi_n^*$.
      Based on the KKT condition \eqref{eq:apdx03:04a}, the optimal beamforming matrix $\bm W_n^*$ can be denoted as $\bm W_n^* = \sum\nolimits_{k=1}^{a_n}\rho_{n, k} \bm\pi_{n, k}\bm\pi_{n, k}^{\H}$ with $\rho_{n, k} \ge 0$.
      In this case, the $n$-th LUE receives no information since the received power of information signal is zero, i.e., $\Tr\br{\bm H_n\bm W_n^*} = \Tr\br{\bm H_n\sum\nolimits_{k=1}^{a_n}\rho_{n, k} \bm\pi_{n, k}\bm\pi_{n, k}^{\H}} = 0$, $n \in \cal N$.
      As a result, the optimal matrix $\bm C_n^*$ satisfies $\Rank\br{\bm C_n^*} = a_n -1$, $n \in \cal N$.
      Moreover, there exists an extra vector $\bm \pi_{n, 0}$ with $\bm \pi_{n, 0}^{\H}\bm C_n^*\bm \pi_{n, 0} = 0$ and $\bm \pi_{n, 0}^{\H}\bm H_n\bm \pi_{n, 0} \neq 0$ such that the optimal beamforming matrix of \eqref{eq:26} is expressed as
      \begin{equation}\label{eq:apdx03:09}
      \bm W_n^* = \sum\limits_{k=0}^{a_n}\rho_{n, k} \bm\pi_{n, k}\bm\pi_{n, k}^{\H}
      \end{equation}
      where $\rho_{n, 0} > 0$ and $\rho_{n, k} \ge 0$, $k = 1, 2, \ldots, a_n$.

      In order to recover the rank-one property of the beamforming matrix, we perform the following manipulations:
      \begin{align}
      \widetilde{\bm W}_n^* =& \bm W_n^* - \sum\limits_{k=1}^{a_n}\rho_{n, k} \bm\pi_{n, k}\bm\pi_{n, k}^{\H}, \forall n \label{eq:apdx03:10a}\\
      \widetilde{\bm Q} =& \bm Q^* + \sumn\sum\limits_{k=1}^{a_n}\rho_{n, k} \bm\pi_{n, k}\bm\pi_{n, k}^{\H} \label{eq:apdx03:10b}
      \end{align}
      where $\{\bm W_n^*\}_{n \in \cal N}$ and $\bm Q^*$ are the optimal solution to \eqref{eq:26}.
      It is straightforward to verify that the beamforming matrices and AN covariance matrix in \eqref{eq:apdx03:10a} and \eqref{eq:apdx03:10b} lie in the feasible region of \eqref{eq:26}, and the corresponding optimal value of \eqref{eq:26} is unchanged.
\end{itemize}

\bibliographystyle{IEEEtran}
\bibliography{dyj_bib}

\end{document}